\documentclass[12pt,a4paper]{article}
\usepackage{amsmath,amssymb,mathrsfs,framed,esint,slashed,graphicx,soul}
\usepackage[colorlinks]{hyperref}
\usepackage{color}
\usepackage[all]{xy}
\usepackage{lipsum}

\newcommand{\rem}[1]{}

\newcommand{\de}{{\rm d}}

\newcommand{\bq}{{\mathbf{q}}}

\newcommand{\bp}{{\mathbf{p}}}

\newcommand{\bZ}{{\mathbf{Z}}}
\newcommand{\bx}{{\boldsymbol{x}}}

\newcommand{\bzeta}{{\boldsymbol{\zeta}}}
\newcommand{\bz}{{\mathbf{z}}}

\newcommand{\bLambda}{{\boldsymbol{\Lambda}}}

\newcommand{\beq}{\begin{equation}}
\newcommand{\eeq}{\end{equation}}
\newcommand{\ben}{\begin{eqnarray}}
\newcommand{\een}{\end{eqnarray}}

\begin{document}

\title{Koopman wavefunctions and\\classical-quantum correlation dynamics}

%\author{Denys I. Bondar$^1$\footnote{\href{dbondar@tulane.edu}}\\
%%\email{dbondar@tulane.edu}
%%\affiliation{
%$^1$Department of Physics and Engineering Physics, Tulane University, New Orleans LA, USA}
%%
%\author{Fran\c{c}ois Gay-Balmaz$^1$\footnote{\href{gaybalma@lmd.ens.fr}}}\\
%%\email{gaybalma@lmd.ens.fr}
%%\affiliation{
%$^1$CNRS and \'Ecole Normale Sup\'erieure, Laboratoire de M\'et\'eorologie Dynamique, Paris, France}
%%
%\author{Cesare Tronci$^1$\footnote{Corresponding author: \href{c.tronci@surrey.ac.uk}}}\\
%%\email[Corresponding author: ]{c.tronci@surrey.ac.uk}
%%\affiliation{
%$^1$Department of Mathematics, University of Surrey, Guildford, UK\\
%%\affiliation{
%$^1$Mathematical Sciences Research Institute,  Berkeley CA, USA\vspace{.3cm}}

\author{Denys I. Bondar$^{1,}$\thanks{\noindent Electronic address: \href{dbondar@tulane.edu}{dbondar@tulane.edu}}\ , Fran\c{c}ois Gay-Balmaz$^{2,}$\thanks{\noindent Electronic address: \href{gaybalma@lmd.ens.fr}{gaybalma@lmd.ens.fr}}\ , Cesare Tronci$^{3,4,}$\thanks{\noindent Corresponding author: \href{c.tronci@surrey.ac.uk}{c.tronci@surrey.ac.uk}} \smallskip 
\\ 
\small
\it $^1$Department of Physics and Engineering Physics, Tulane University, New Orleans LA, USA
\\ 
\small
\it $^2$CNRS and \'Ecole Normale Sup\'erieure, Laboratoire de M\'et\'eorologie Dynamique, Paris, France
\\
\small
\it $^3$Department of Mathematics, University of Surrey, Guildford, United Kingdom
\\
\small
\it 
$^4$Mathematical Sciences Research Institute,  Berkeley CA, USA}
\date{$\,$}

\date{$\,$}

\maketitle

\begin{abstract} 
Upon revisiting the Hamiltonian structure of  classical wavefunctions in Koopman-von Neumann theory, this paper addresses the long-standing problem of formulating a dynamical theory of  classical-quantum coupling. The proposed model not only describes the influence of a classical system onto a quantum one, but also the reverse effect -- the quantum backreaction. These interactions are described by a new Hamiltonian wave equation overcoming shortcomings of currently employed models. For example, the density matrix of the quantum subsystem is always positive-definite. While the Liouville density of the classical subsystem is generally allowed to be unsigned, its sign is shown to be preserved in time for a specific infinite family of hybrid classical-quantum systems. The proposed  description is illustrated and compared with previous theories using the exactly solvable model of a degenerate two-level quantum system coupled to a classical harmonic oscillator. 
\end{abstract}

\bigskip

%\section{Significance statement}
%Since the dawn of quantum mechanics a Hamiltonian description for the interaction between quantum and classical systems has been sought, however with limited success. We finally furnish such a model that overcomes shortcomings of previous attempts. This finding may open new venues in the foundation of quantum mechanics as well as enable the development new generation computational methods in quantum chemistry.

\section{Introduction}
Classical-quantum coupling has been an open problem since the rise of quantum mechanics. Bohr's concept of uncontrollable disturbance \cite{Bohr1935} affecting both classical and quantum systems during the measurement process has attracted much attention over the decades and it would be unfeasible to provide here the enormous list of works in this field. The effect of the uncontrollable disturbance on the quantum system is often known under the name of `decoherence' \cite{Zurek} and it manifests in terms of non-unitary dynamics and purity non-preservation \cite{VonNeumann}. Recently the dynamics of a classical measuring device interacting with a quantum system has become a subject of experimental investigations (see, e.g., \cite{HacohenGourgy}). Over the last four decades, the apparent impossibility of a fully deterministic Hamiltonian description of classical-quantum coupling has been overcome by modeling decoherence in terms of Markov stochastic processes. Then,  the quantum Lindblad equation  \cite{Lindblad,GoKoSu} has emerged as the most general type of a Markovian master equation describing the evolution of a positive-definite and unit-trace quantum density matrix. Lindblad's theory, however, does not comprise the dynamics of the classical subsystem, which is simply treated as a thermodynamical bath.  

In many physical contexts (in e.g., quantum chemistry and laser cooling), the systems under consideration are to be modeled as hybrid evolution to capture the coupling between  electronic degrees of freedom and heavy nuclei. Then, it becomes essential to capture the `quantum backreaction' -- the quantum feedback force on the evolution of the classical system (i.e., the nuclei). To this purpose, in 1981 Aleksandrov \cite{Aleksandrov} and Gerasimenko \cite{Gerasimenko} independently proposed the following quantum-classical {Liouville} equation for an operator-valued density on phase-space $\widehat{\cal D}(\bq,\bp,t)$:
\beq\label{Aleq}
\frac{\partial\widehat{\cal D}}{\partial t}=-i\hbar^{-1}[\widehat{H},\widehat{\cal D}]+\frac12\left(\{\widehat{H},\widehat{\cal D}\}-\{\widehat{\cal D},\widehat{H}\}\right),
\eeq
where $\widehat{H}(\bq,\bp)$ is the operator-valued Hamiltonian function  and we have used the standard notation for commutators $[\, , \,]$ and canonical Poisson brackets $\{\,, \,\}$. The work by Aleksandrov and Gerasimenko has been highly influential and its Wigner-transformed variant is currently used for modeling purposes \cite{Kapral}. Shortly after its appearance, the Aleksandrov-Gerasimenko (AG) equation \eqref{Aleq} was rediscovered in \cite{boucher}, where it was derived from first principles in terms of invariance properties under canonical and unitary transformations. However, although  equation \eqref{Aleq} conserves the total energy $h=\operatorname{\sf Tr}\!\int\!\widehat{H}\widehat{\cal D}\,\de^{3}q\,\de^3p$, the quantum density matrix $\int\!\widehat{\cal D}\,\de^{3}q\,\de^3p$ is not positive definite. More importantly, the AG equation  lacks a Hamiltonian structure and this is due to the fact that the binary operation  in the right-hand side of \eqref{Aleq}  does not satisfy the Jacobi identity and thus it is not a type of Poisson bracket \cite{CaroSalcedo,Salcedo,Prezhdo}. 
 In this case, the absence of a Hamiltonian structure leads to time-irreversible dynamics \cite{Sergi}, thereby indicating a possible entropy production, which is normally formulated as an H-theorem. However, entropy-preserving dynamics requires the formulation of time-reversible models possessing a Hamiltonian structure, which is indeed available in the case of isolated classical and quantum systems. Then, a Hamiltonian model of  quantum-classical hybrid dynamics becomes necessary to model recurrent evolution such as Rabi oscillations.
%Since the isolated classical and quantum dynamics both have Hamiltonian descriptions, it is natural to expect that the classical-quantum hybrid should inherit the same structure. 
Despite several efforts \cite{Anderson,Marmo,Diosi,Elze,HaRe,PrKi,Buric,Sahoo,Shirokov},  Lie-algebraic arguments \cite{CaroSalcedo,Salcedo} tend to exclude the existence of a closed equation for $\widehat{\cal D}$ possessing a Hamiltonian structure (i.e. comprising the Jacobi identity). 

Another stream of research on classical-quantum coupling goes back to Sudarshan's measurement theory  \cite{Sudarshan} of 1976. Therein,   Sudarshan proposed to couple classical and quantum dynamics by exploiting the Koopman-von Neumann (KvN) formulation of classical dynamics in terms of classical wavefunctions \cite{Koopman,VonNeumann2}. Rediscovered in several instances \cite{Wiener,tHooft}, this reformulation of classical mechanics has been attracting increasing attention \cite{Bondar,Ghose,Mauro, Klein,Wilczek,Ramos,Viennot}. See also \cite{Mezic} for a broad review of general applications of Koopman operators.  In the KvN construction, the classical Liouville density $\rho(\bq,\bp,t)$ is expressed as $\rho=|\Psi|^2$, where $\Psi(\bq,\bp,t)$ is a wavefunction obeying the KvN equation
\beq\label{KvNeq}
i\hbar\frac{\partial \Psi}{\partial t}=\{i\hbar H,\Psi\}=:\widehat{L}_H\Psi
\,.
\eeq
Here, we have introduced the Hermitian Liouvillian operator $\widehat{L}_H\cdot=i\hbar\{ H,\cdot\}$. A direct verification shows that the prescription $\rho=|\Psi|^2$ returns the classical Liouville equation $\partial_t\rho=\{H,\rho\}$. Upon working in the Heisenberg picture, Sudarshan extended the above equation \eqref{KvNeq} to include the interaction with quantum degrees of freedom by invoking special superselection rules to enforce physical consistency \cite{Sudarshan}. Although extremely  inspiring, this approach has received some criticism over the years \cite{Barcelo,PeTe,Terno,Sudarshan2} mainly because the role of the superselection rules remains somewhat unclear. Still, one of the advantages of Sudarshan's proposal is that Koopman wavefunctions possess a simple canonical Hamiltonian structure formally equivalent to that underlying Schr\"odinger's equation. Indeed, this feature provides a great simplification  over the AG approach, which instead is based on density operators and  Wigner functions both carrying highly noncanonical Lie-Poisson brackets \cite{BiMo}.

While several hybrid theories appearing in the literature may offer good approximations of classical-quantum coupling, a Hamiltonian theory is still lacking and this poses specific problems concerning consistent transformation properties. This paper addresses this problem by  following up on Sudarshan's idea of exploiting classical wavefunctions. Upon combining this approach with Hamiltonian methods, we shall show that KvN theory can be easily extended in such a way that its Hamiltonian functional coincides with the  physical energy. In the second part of the paper we shall infer a Hamiltonian theory for classical-quantum coupling by using the extended KvN representation within the context of geometric quantization. The proposed classical-quantum wave equation is illustrated on the exactly solvable model of a degenerate two-level quantum system quadratically coupled to a one-dimensional classical harmonic oscillator.

\section{Koopman wavefunctions\label{Sec:KvH}}

We begin by looking at the Hamiltonian structure of the KvN equation \eqref{KvNeq}. This structure is particularly transparent when looking at its variational formulation
\beq\label{DF-KvN}
\delta\int_{t_1}^{t_2}\!
\!\int\!\Big(\hbar\operatorname{\sf Re}(i\Psi^*\partial_t\Psi)-\Psi^*\widehat{L}_H\Psi\Big)\,\de^6z\,\de t
%\langle\Psi,i\hbar\partial_t\Psi-\widehat{L}_H\Psi\rangle\,\de t
=0
\,,
\eeq
%where we have denoted $\langle\Psi_1,\Psi_2\rangle=\operatorname{\sf Re}\langle\Psi_1|\Psi_2\rangle$ and $\langle\cdot|\cdot\rangle$ is the standard inner product. 
which leads to a few observations. 

First, the Hamiltonian functional for the KvN equation \eqref{KvNeq} is written as
$h(\Psi)=\int\!\Psi^*\widehat{L}_H\Psi\,\de^6z\\=\hbar\int\!H\operatorname{\sf Im}\{\Psi^*,\Psi\}\,\de^6 z$, 
where we have denoted $\bz=(\bq,\bp)$. Then, we observe that the Hamiltonian functional for the KvN equation
does not coincide with the total physical energy, which instead would read $\int\!H|\Psi|^2\,\de^6 z$  (according to the  prescription $\rho=|\Psi|^2$). 

The second observation is that the quantity $\operatorname{\sf Im}\{\Psi^*,\Psi\}$ satisfies the classical Liouville equation and thus, in principle we could set $\rho=\operatorname{\sf Im}\{\Psi^*,\Psi\}$. Borrowing a terminology from fluid dynamics \cite{Clebsch}, this expression is often known as a \emph{Clebsch representation} \cite{HoKu,MaWe,Morrison} in the context of Geometric Mechanics \cite{MaRa,HoScSt}. However, here we are left with the insurmountable  problem that $\int \!\operatorname{\sf Im}\{\Psi^*,\Psi\}\,\de^6z=0$.

The third observation is more fundamental: we remark that the KvN Lagrangian (the integrand in \eqref{DF-KvN}) {is not covariant with respect to local phase transformations $\Psi(\bz)\mapsto e^{i\varphi(\bz)}\Psi(\bz)$}. However, this particular problem can be overcome by using the minimal coupling method in gauge theory. Let us introduce the multiplicative operator $\widehat{\bZ}=\bz$ and its canonical conjugate $\widehat{\bLambda}=-i\hbar\nabla$, and let us rewrite the Liouvillian as $\widehat{L}_H=\mathbf{X}_H(\widehat{\bZ})\cdot\widehat{\bLambda}$. Here,  $\mathbf{X}_H=J\nabla H$ is the classical Hamiltonian vector field and 
\[
J= \left( \begin{matrix}
\boldsymbol{0} & \boldsymbol{1}\\
-\boldsymbol{1} & \boldsymbol{0}
\end{matrix}
\right),
\]
so that  $[\widehat{Z}^i,\widehat{\Lambda}^j]=i\hbar \delta^{ij}$.
 Then, if $(\Phi,\boldsymbol{\cal A})$ are the components of a $U(1)-$gauge potential, a gauge-covariant {Liouvillian} is constructed by the replacement 
%\beq
%%\mathbf{X}_H(\widehat{\bZ})\cdot\widehat{\bLambda}=
%%\widehat{L}_H\to\widehat{\cal L}_H
%%,\quad\!\! \text{with}\quad\!
%%\widehat{\cal L}_H:=
%\mathbf{X}_H(\widehat{\bZ})\cdot\widehat{\bLambda}
%\ \longrightarrow\ 
%\Phi(\widehat{\bZ})+\mathbf{X}_H(\widehat{\bZ})\cdot(\widehat{\bLambda}-\boldsymbol{\cal A}(\widehat{\bZ}))
%,
%\label{repl}
%\eeq
{
\beq\label{repl}
i\hbar\partial_t\longrightarrow i\hbar\partial_t-\Phi
\,,\qquad\qquad
i\hbar\nabla\longrightarrow i\hbar\nabla+\boldsymbol{\cal A}
\,.
\eeq
Then, the covariant Liouvillian takes the form
\beq\label{cov_liouv}
\widehat{\cal L}_H:=\Phi(\widehat{\bZ})+\mathbf{X}_H(\widehat{\bZ})\cdot(\widehat{\bLambda}-\boldsymbol{\cal A}(\widehat{\bZ}))
\,.
\eeq
Now}, the choice of gauge potential is usually prescribed in prequantization theory \cite{GBTr,Souriau,VanHove} as follows:
\beq\label{Gauconn}
\Phi(\bz)=H(\bz)
\,,\qquad\ 
\boldsymbol{\cal A}(\bz)\cdot\de\bz=\bp\cdot\de\bq
%=-\frac12J\bz\cdot\de\bz%%this is correct with the last change in sign
\,.
\eeq
{Here, the differential form $\boldsymbol{\cal A}(\bz)\cdot\de\bz$ is known as the \emph{symplectic potential}, so that the standard symplectic form is obtained as $\omega=-\de \boldsymbol{\cal A}$, or equivalently $\nabla\boldsymbol{\cal A}-(\nabla\boldsymbol{\cal A})^T=-J$.}
Under the replacement \eqref{repl},  the variational principle \eqref{DF-KvN} yields the modified KvN equation
\beq\label{modKvN}
i\hbar\frac{\partial \Psi}{\partial t}=\{i\hbar H,\Psi\}-
{\bigg(\bp\cdot\frac{\partial H}{\partial \bp} -H\bigg)\Psi}
%\Big(\frac12\bz\cdot\nabla H-H\Big)\Psi
%=:\widehat{\cal L}_H\Psi
\,.
\eeq
{
First formulated in 1972 by Kostant \cite{Kostant1}, this equation has appeared in a few works \cite{boucher,Gunther,Sugny,Klein}, where it was noted that the expression $\rho=|\Psi|^2$ again satisfies the classical Liouville equation. In addition, we emphasize that the phase term in equation \eqref{modKvN} is readily seen to coincide with the Lagrangian 
\[
\mathscr{L}=\bp\cdot\partial_{\bp}H-H,
\] 
thereby reminding the important relation between phases and Lagrangians going back to Feynman's thesis \cite{Feynman}. The relation between the Lagrangian and the classical phase is made explicit by replacing the polar form $\Psi=\sqrt{D}e^{iS/\hbar}$ 
in \eqref{modKvN}, thereby obtaining
\[
\frac{\partial D}{\partial t}+\{ D,H\}=0
\,,\qquad\quad
\frac{\partial S}{\partial t}+\{ S,H\}=\mathscr{L}
.
\]
Then, we recognize that while KvN theory is totally equivalent to the classical Liouville equation, equation \eqref{modKvN} also carries information about the classical phase.
Recently, the crucial role of both classical and quantum phases was also exploited in connection to the Hamilton-Jacobi theory \cite{Scully1,Scully2}, although in that context the wavefunction is defined only on the position space.

As a further remark, we notice that different gauge fixings are possible in alternative to \eqref{Gauconn}. For example, the \emph{harmonic oscillator gauge} 
\beq
\boldsymbol{\cal A}\cdot\de\bz=\frac12J\bz\cdot\de\bz= \frac12 \left( \bp\cdot\de\bq - \bq\cdot\de\bp\right)
\label{HOg}
\eeq
used in \cite{GBTr,Sugny} is convenient for homogeneous quadratic Hamiltonians as in this case the corresponding phase term $\Phi-\mathbf{X}_H\cdot\boldsymbol{\cal A}=H-{\bz\cdot\nabla H/2}$ vanishes identically. Moreover, since $\bp\cdot\de\bq$ is also known as the ``Liouville one-form'', we shall refer to the gauge in \eqref{Gauconn} as the \emph{Liouville gauge}. Both gauges will be used in this paper, depending on convenience.}

First appeared in van Hove's prequantization theory \cite{VanHove}, the {covariant} Liouvillian $\widehat{\cal L}_H$ is known as a \emph{prequantum operator} \cite{Hall} and it satisfies the Lie algebra relation $[\widehat{\cal L}_H,\widehat{\cal L}_K]=i\hbar\widehat{\cal L}_{\{H,K\}}$. In addition, we have a one-to-one correspondence between the Hamiltonian $H$ and the Hermitian operator $\widehat{\cal L}_H$ (unlike the correspondence $H\mapsto\widehat{L}_H$, which is many-to-one). 
In the Heisenberg picture (here denoted by the superscript $\sf H$), {one defines ${\widehat{\cal L}}_A^{\,\sf H}(t):=U(t)^\dagger{\widehat{\cal L}}_A U(t)$ where $U(t)=\exp({-i{\widehat{\cal L}}_H}t/\hbar)$ is the classical propagator for a given Hamiltonian $H$. Then, this yields  $\de{\widehat{\cal L}}_A^{\,\sf H}/\de t=i\hbar^{-1}[\widehat{\cal L}_H,\widehat{\cal L}_A^{\,\sf H}]$ as well as $\widehat{\cal L}_H^{\,\sf H}=\widehat{\cal L}_H$. By construction, one has the general property  $\widehat{\cal L}_A^{\,\sf H}= \widehat{\cal L}_{A^{\sf H}}$, where $A^{\sf H}(t)=\exp({i{\widehat{L}}_H}t/\hbar)A$ and the Liouvillian $\widehat{L}_H$ is given as in \eqref{KvNeq}.  See Appendix \ref{Appendix_A} for further explanations. 
Therefore, the  Heisenberg equation for $\widehat{\cal L}_A^{\,\sf H}$ implies the usual dynamics $\de{A}^{\sf H}/\de t=\{A^{\sf H},H\}$ for classical observables}. 

Partly inspired by Kirillov \cite{Kirillov}, here we shall call  \eqref{modKvN} the \emph{Koopman-van Hove (KvH) equation} and  address the reader also to \cite{Sugny,Klein} for more discussions on how prequantization relates to KvN theory.
Let us now examine the Hamiltonian structure of the modified KvN equation \eqref{modKvN}. The variational principle $\delta\int_{t_1}^{t_2}
\!\!\int\!\big(\hbar\operatorname{\sf Re}(i\Psi^*\partial_t\Psi)-\Psi^*\widehat{\cal L}_H\Psi\big)\,\de^6z\,\de t=0$ determines the Hamiltonian functional
\beq
h=
%\langle\Psi|\widehat{\cal L}_H\Psi\rangle=
\int\!\Psi^*\widehat{\cal L}_H\Psi\,\de^6z=
\int\!H\big(|\Psi|^2+\operatorname{\sf div}\boldsymbol{\cal J}\big)\,\de^6 z
\,,
\label{KvNenergy}
\eeq
with
\[
\boldsymbol{\cal J}=\Psi^*\,\widehat{\!\boldsymbol{\cal Z}\,}_{\!\!+}\Psi
\,,\qquad\text{and}\qquad
\widehat{\!\boldsymbol{\cal Z}\,}_{\!\!\pm}:={J}(\pm\widehat{\bLambda}-\boldsymbol{\cal A})
%\frac12{\widehat{\bf Z}}\pm J\widehat{\bLambda}
\,.
\]
We note in passing that the operators $\widehat{\!\boldsymbol{\cal Z}\,}_{\!\!\pm}$ satisfy the  commutation relations $[\,\widehat{\!{\cal Z}\,}_{\!\!\pm}^i,\,\widehat{\!{\cal Z}\,}_{\!\!\pm}^j]=\mp i\hbar J^{ij}$ and $[\,\widehat{\!{\cal Z}\,}_{\!\!\pm}^i,\,\widehat{\!{\cal Z}\,}_{\!\mp}^j]=0$, which were used in \cite{deGosson,Bondar} {(by adopting the harmonic oscillator gauge \eqref{HOg})} to rewrite quantum theory in terms of wavefunctions on phase-space. From equation \eqref{KvNenergy}, we see that the quantity $|\Psi|^2+\operatorname{\sf div}\boldsymbol{\cal J}$ emerges as an alternative Clebsch representation for the Liouville density. More specifically, this quantity is a momentum map \cite{Sternberg,Sternberg2,MaRa,HoScSt} for the  group of strict contact transformations generated by the operator $i\hbar^{-1}\widehat{\cal L}_H$ \cite{VanHove}, where
\beq
\widehat{\cal L}_H={H}-\nabla{H}\cdot\,\widehat{\!\boldsymbol{\cal Z}\,}_{\!\!+}
\,.
\label{CovL}
\eeq
While some of this material is illustrated in the Appendix, we shall leave a more thorough discussion of these aspects for future work. Here we emphasize that the momentum map property enforces the quantity $|\Psi|^2+\operatorname{\sf div}\boldsymbol{\cal J}$ to satisfy the classical Liouville equation, as it can be verified by a direct and lengthy calculation.

At this point, given the expression of the total energy \eqref{KvNenergy}, we insist that this must be equal to the total physical energy $\int\!H\rho\,\de^6z$, and thus we are led to the identification
\begin{align}\nonumber
\rho=&\ |\Psi|^2+\operatorname{\sf div}\!\big(\Psi^*\widehat{\!\boldsymbol{\cal Z}\,}_{\!\!+}\Psi\big)\\
=&\ 
|\Psi|^2-\operatorname{\sf div}(J\boldsymbol{\cal A}\,|\Psi|^2)
+\hbar
\operatorname{\sf Im}\{\Psi^*,\Psi\}
\label{Clebsch}
\,.
\end{align}
Although we observe that this expression for the Liouville density is not positive-definite, its sign is preserved in time since the Liouville equation is a characteristic equation. Remarkably, we notice that the term $\operatorname{\sf div}\boldsymbol{\cal J}$ does not contribute to the total probability, so that $\int\!\rho\,\de^6z=\int|\Psi|^2\,\de^6z=1$. On the other hand, the same divergence term does contribute to expectation values, so that e.g.
$
\langle\bz\rangle=\int\!\bz\rho\,\de^6z=\int\!\Psi^*\widehat{\!\boldsymbol{\cal Z}\,}_{\!\!-}\Psi\,\de^6z
$.
As shown in \cite{Bondar} {by adopting the harmonic oscillator gauge \eqref{HOg}}, this last relation returns the usual Ehrenfest equations for the expectation dynamics of  canonical observables. 
%Another feature of  relation \eqref{Clebsch} is that the presence of first-order derivatives may allow for $\delta-$like singularities (point particle solutions), which  instead are removed by the usual KvN prescription $\rho=|\Psi|^2$. 

Lastly, we remark that the entire discussion can be repeated by replacing classical wavefunctions with {(possibly unsigned)} density-like operators  mimicking Von Neumann's density matrix \cite{boucher}. Then, equation \eqref{modKvN} is recovered upon setting $\widehat{D}(\bz,\bz',t)=\Psi(\bz,t)\Psi^*(\bz',t)$ in the evolution equation  
$
i\hbar\partial_t\widehat{D}=[\widehat{\cal L}_H,\widehat{D}]
%\label{D-eq}
$.
%In passing, we notice that equation \eqref{D-eq} allows for mixture-like solutions of the type $\widehat{D}(\bz,\bz')=\sum_kw_k\Psi_k(\bz,t)\Psi_k^*(\bz',t)$, which can be used for statistical sampling purposes so that $\rho=\sum_k w_k \rho_k$. These solutions mimic the concept of quantum mixtures, which in turn are another type of momentum map whose generalizations appear in different models for nonadiabatic molecular dynamics \cite{Tronci18}.
%In this Section we have extended the KvN theory to include gauge-covariance under local phases and we have redefined the Liouville density in such a way that the KvN energy equals the total physical energy. 
In the following sections, we shall further extend the present gauge-covariant KvH construction  to include the coupling to  quantum degrees of freedom.

\section{Hybrid classical-quantum dynamics}

The formulation of hybrid classical-quantum dynamics is usually based on fully quantum treatments, in which some kind of factorization ansatz is invoked on the wave function. This ansatz is then followed by a classical limit on the factor that is meant to model the classical particle. 

Here, we propose a different perspective: we shall start with the KvH construction for two classical particles and we shall perform a formal quantization procedure on one of them. {This can be achieved in different ways, depending on the particular quantization procedure. For example, Gerasimenko proposed a similar approach in the context of Weyl quantization  \cite{Gerasimenko}, while the KvH equation \eqref{modKvN} was formulated by Kostant \cite{Kostant1} in the context of geometric quantization \cite{Hall,Kostant}. Here, however, we shall adopt a simpler approach which consists in a partial canonical quantization on the 2-particle Hamiltonian. We consider the KvH equation \eqref{modKvN} for a wavefunction $\Psi(\bz,\bzeta)$ representing  two particles with coordinates $\bz=(\bq,\bp)$ and $\bzeta=(\boldsymbol{x},\boldsymbol{\mu})$, and fix a Hamiltonian $H(\bz,\bzeta)$. Then, we apply canonical quantization only to the coordinates $(\boldsymbol{x},\boldsymbol{\mu})$, so that one replaces $\boldsymbol{x}\rightarrow\widehat{\boldsymbol{x}}$ (quantum position operator) and $\boldsymbol{\mu}\rightarrow\widehat{\boldsymbol{p}}:=-i\hbar\partial_{\boldsymbol{x}}$ (quantum momentum operator) in the 2-particle Hamiltonian $H$, which thus becomes an operator-valued function $\widehat{H}(\bz,\widehat{\boldsymbol{x}},\widehat{\boldsymbol{p}})$ and the coordinate $\boldsymbol{\mu}$ has been eliminated. The hybrid Hamiltonian is then replaced in \eqref{CovL} to obtain the hybrid Liouvillian $\widehat{\cal L}_{\widehat{H}}=\widehat{H}-\nabla\widehat{H}\cdot\,\widehat{\!\boldsymbol{\cal Z}\,}_{\!\!+}$, with $\,\widehat{\!\boldsymbol{\cal Z}\,}_{\!\!+}=-{J}(i\hbar\nabla_{\!\bz}+\boldsymbol{\cal A}(\bz))$.} Eventually, one is left with the following \emph{classical-quantum wave equation} for the hybrid wavefunction (here, denoted by $\Upsilon(\bz,\boldsymbol{x})$):
\beq\label{UpsEq}
i\hbar\partial_{t\!}\Upsilon=
\widehat{H}\Upsilon-\nabla\widehat{H}\cdot\,\widehat{\!\boldsymbol{\cal Z}\,}_{\!\!+}\Upsilon
=:\widehat{\cal L}_{\widehat{H}}\Upsilon\,.
\eeq
{For example, performing the partial quantization on the 2-particle Hamiltonian $H(\bz,\bzeta)=p^2/2M+\mu^2/2m+V(\bq,\boldsymbol{x})$ yields the hybrid classical-quantum Hamiltonian}
\beq
\widehat{H}=-\frac{\hbar^2}{2m}\Delta_\bx+\frac1{2M}p^2+V(\bq,\boldsymbol{x})
\,.
\label{HybridHamiltonian}
\eeq
%and we have defined the hybrid Liouvillian  $\widehat{\cal L}_{\widehat{H}}=\widehat{H}-\nabla\widehat{H}\cdot\,\widehat{\!\boldsymbol{\cal Z}\,}_{\!\!+}$.
 
 Equations with a similar structure to \eqref{UpsEq} were shown to occur in the Hamiltonian dynamics of quantum expectation values \cite{BLTr1,BLTr2}. {Equations similar to 
 \eqref{UpsEq} were also obtained in \cite{boucher} upon discarding the phase terms in the KvH equation \eqref{modKvN}, that is by considering the standard KvN equation \eqref{KvNeq}. In that paper the authors  rejected  their equations because of interpretative issues and the absence of a conserved a positive energy.} Here, we  point out  that, since $\widehat{\cal L}_{\widehat{H}}$ is Hermitian, then \eqref{UpsEq} is actually a Hamiltonian equation possessing a variational principle of the type 
\beq
\delta\int_{t_1}^{t_2}\!\operatorname{\sf Re}\big\langle\Upsilon\big|\big(i\hbar\partial_t-\widehat{\cal L}_{\widehat{H}}\big)\Upsilon\big\rangle\,\de t=0
\,,
\label{hybridVP}
\eeq
%\begin{multline}
%\delta\int_{t_1}^{t_2}\!\!\operatorname{\sf Tr}\!\bigg[\int\!\bigg(\hbar\operatorname{\sf Re}\big(i\Upsilon^\dagger(\bz)\partial_{t\!}\Upsilon(\bz)\big)
%\\
%-\Upsilon^\dagger(\bz)\widehat{\cal L}_{\widehat{H}}\Upsilon(\bz)\bigg)\, \de^6z\bigg]\,\de t=0
%\,,
%\end{multline}
thereby preserving the  the energy invariant
\beq
h=\langle\Upsilon|\widehat{\cal L}_{\widehat{H}}\Upsilon\rangle=\operatorname{\sf Tr}\!\int\!\Upsilon^\dagger(\bz)\,\widehat{\cal L}_{\widehat{H}}\Upsilon(\bz)\,\de^6z
\,.
\label{TotEn}
\eeq
Here, the dagger symbol denotes the adjoint in  the quantum coordinates and similarly for the trace, so that  $\langle\Upsilon_1|\Upsilon_2\rangle=\operatorname{\sf Tr}\int\Upsilon_1^\dagger(\bz)\Upsilon_2(\bz)\,\de^6z$.

Now we construct a generalized density operator $\widehat{\cal D}$ so that the total energy \eqref{TotEn} reads $h=\operatorname{\sf Tr}\!\int\!\widehat{H}\widehat{\cal D}\,\de^6z$. Actually, the latter relation is obtained by a direct manipulation of the expression \eqref{TotEn}, upon defining
%\beq
%h=\operatorname{\sf Tr}\!\int\!\widehat{H}\widehat{\cal D}\,\de^6z
%\eeq
%with
\beq\label{Ddef}
\begin{aligned}
\widehat{\cal D}(\bz)&=\Upsilon(\bz)\Upsilon^\dagger(\bz)+\operatorname{\sf div}\!\big(\Upsilon(\bz)\,\widehat{\!\boldsymbol{\cal Z}\,}_{\!\!-}\Upsilon^\dagger(\bz)\big)\\
&={\Upsilon(\bz)\Upsilon^\dagger(\bz) - \operatorname{div}\big(J\boldsymbol{\cal A} \Upsilon(\bz) \Upsilon^\dagger(\bz)\big) + {\rm i}\hbar\{\Upsilon(\bz),\Upsilon^\dagger(\bz)\}}\,.
\end{aligned}
\eeq
This quantity plays the role of the AG generalized density in \eqref{Aleq} and it belongs to the dual of the tensor product space of phase-space functions and Hermitian operators on the quantum state space. Since the latter tensor space is not a Lie algebra (notice $[\widehat{\cal L}_{\widehat{F}},\widehat{\cal L}_{\widehat{G}}]\neq\widehat{\cal L}_{\widehat{K}}$ for some $\widehat{K}(\bz)$), $\widehat{\cal D}$ does not carry a standard momentum map structure and thus it cannot possess a closed Hamiltonian equation, in agreement with \cite{CaroSalcedo,Salcedo}. 

In addition, we remark that $\widehat{\cal D}$ is generally not positive definite and, unlike the purely classical case, its sign is not preserved in time. This  feature (also occurring in the AG equation \eqref{Aleq}) was justified in \cite{boucher} by analogies with Wigner quasi-probability densities. In the present context, the quantum density matrix and the classical Liouville density read
\begin{align}
&\hat{\rho}=\!\int\!\widehat{\cal D}(\bz)\,\de^6z=\!\int\!\Upsilon(\bz)\Upsilon^\dagger(\bz)\,\de^6z \label{EqQuantumDensity}
\,
\\
&\rho(\bz)= \!\operatorname{\sf Tr}\widehat{\cal D}(\bz)=\!\operatorname{\sf Tr}\!\Big[\Upsilon(\bz)\Upsilon^\dagger(\bz)\!+\!\operatorname{\sf div}\!\big(\Upsilon(\bz)\,\widehat{\!\boldsymbol{\cal Z}\,}_{\!\!-\,}\Upsilon^\dagger(\bz)\big)\Big]. \label{EqClassicalDensity}
\end{align}
Then, while the quantum density matrix is positive definite by construction (unlike the AG theory \cite{Aleksandrov,Gerasimenko}),  the classical Liouville density  is allowed to become negative in the general case of  classical-quantum interaction. 

A further consequence of equation \eqref{UpsEq} is obtained by simply applying Ehrenfest's theorem: indeed, the latter yields the following expectation value equation for quantum-classical observables $\widehat{A}(\bz)$:
\beq
i\hbar\frac{\de \langle\widehat{A}\rangle}{\de t}=\big\langle\Upsilon\,\big|\big[\widehat{\cal L}_{\widehat{A}},\widehat{\cal L}_{\widehat{H}}\big]\Upsilon\big\rangle
\,,
\label{QCExp}
\eeq
where we have defined $ \langle\widehat{A}\rangle=\operatorname{\sf Tr}\!\int\!\widehat{A}\widehat{\cal D}\,\de^6z=\langle\Upsilon|\widehat{\cal L}_{\widehat{A}}\Upsilon\rangle$. Then, the usual conservation laws are recovered in the case $\big[\widehat{\cal L}_{\widehat{A}},\widehat{\cal L}_{\widehat{H}}\big]=0$. For example, {upon denoting $\widehat{\boldsymbol{p}}=-i\hbar\nabla_{\!\boldsymbol{x}}$, the case $\widehat{\!\boldsymbol{A}\,}=\bp+\widehat{\boldsymbol{p}}$  recovers momentum conservation   whenever the generic Hamiltonian \eqref{HybridHamiltonian} involves a translation-invariant potential $V(\bq-\widehat{\bx})$. (Here, $\widehat{\bx}$ denotes the quantum position operator). Indeed, the conservation of  total momentum $[\widehat{\cal L}_{\bp+\widehat{\boldsymbol{p}}},\widehat{\cal L}_{\widehat{H}}]=0$ follows from the relations $[\widehat{\cal L}_{\bp},\widehat{\cal L}_{\widehat{H}}]=i\hbar\widehat{\cal L}_{\{\bp,V\}}$ and $[\widehat{\cal L}_{\widehat{\boldsymbol{p}}},\widehat{\cal L}_{\widehat{H}}]=\widehat{\cal L}_{\,[\widehat{\boldsymbol{p}},V]}$, since we have ${i\hbar{\{\bp,V\}}+[\widehat{\boldsymbol{p}},V]=0}$.} We remark that the  expectation dynamics \eqref{QCExp} differs from the corresponding result obtained from the AG equation \eqref{Aleq}.

We conclude by presenting the dynamics of $\widehat{\cal D}$. As we pointed out, $\widehat{\cal D}$ does not possess a closed Hamiltonian equation: this means that its evolution can only be expressed in terms of $\Upsilon$. In the case of a finite-dimensional quantum state space, a lengthy computation shows that (in index notation)
\begin{align}\nonumber
\partial_t{\widehat{\cal D}}_{\alpha\beta}=&\
-i\hbar^{-1}\big[\widehat{H},\widehat{\cal D}\big]_{\alpha\beta}
%\\
%&\ 
+\big\{
\widehat{H},\widehat{\cal D}
\big\}_{\alpha\beta}
-
\big\{
\widehat{\cal D},\widehat{H}
\big\}_{\alpha\beta}
\\\nonumber
&\ +
\big\{J\boldsymbol{\cal A}\Upsilon\Upsilon^\dagger,\nabla\widehat{H}\big\}_{\alpha\beta}
-
\big\{\nabla\widehat{H},J\boldsymbol{\cal A}\Upsilon\Upsilon^\dagger\big\}_{\alpha\beta}
\\\nonumber
&\ 
+i\hbar^{-1}\operatorname{\sf div}\!\big[J\boldsymbol{\cal A}\cdot\nabla \widehat{H},J\boldsymbol{\cal A}\Upsilon\Upsilon^\dagger\big]_{\alpha\beta}
+\big[J\boldsymbol{\cal A} \cdot\nabla \widehat{H},\{\Upsilon,\Upsilon^\dagger\}\big]_{\alpha\beta}
\\\nonumber
&\ 
 +
\operatorname{\sf div}\big(
\big\{\widehat{H}_{\alpha\gamma},J\boldsymbol{\cal A}\Upsilon_\beta^*\big\}\Upsilon_\gamma
-
\big\{J\boldsymbol{\cal A}\Upsilon_\alpha,\widehat{H}_{\gamma\beta}\big\}\Upsilon_\gamma^*
\big)
\\\nonumber
&\ 
+\Upsilon_\gamma\big\{J\boldsymbol{\cal A}\cdot\nabla \widehat{H}_{\alpha\gamma},\Upsilon_\beta^*\big\}
-
\big\{
\Upsilon_\alpha,
J\boldsymbol{\cal A}\cdot\nabla \widehat{H}_{\gamma_\beta}
\big\}\Upsilon_\gamma^*
\\
&\ 
-i\hbar\{\Upsilon_\gamma,\{\widehat{H}_{\alpha\gamma},\Upsilon^*_{\beta}\}\}
+
i\hbar\{\{\Upsilon_\alpha,\widehat{H}_{\gamma\beta}\},\Upsilon^*_\gamma\}
\,.
\label{Deqn}
\end{align}
where all quantities are evaluated at $\bz$. Despite the striking similarity between the first line above and the AG equation \eqref{Aleq}, the remaining terms in the $\widehat{\cal D}-$equation show  that the classical-quantum interaction may be more involved than one might have expected. Nevertheless, the intricate nature of classical-quantum coupling becomes hidden by the formal simplicity of the following equations for the quantum and classical densities:
\beq
\label{PartialTraces}
i\hbar\partial_t\hat{\rho}=\int[\widehat{H},\widehat{\cal D}]\,\de^6z
\,,\qquad\ 
\partial_t\rho=\operatorname{\sf Tr}\{\widehat{H},\widehat{\cal D}\}
\,,
\eeq
which coincide formally with the corresponding result obtained by using the AG equation \eqref{Aleq}. {We notice, however, that the AG theory is fundamentally different from the classical-quantum model formulated here. As we already mentioned, the AG equation is not Hamiltonian and it does not generally preserve the positivity of the quantum density matrix $\hat{\rho}=\!\int\!\widehat{\cal D}(\bz)\,\de^6z$. In addition, the classical-quantum wave equation \eqref{UpsEq} represents a significant simplification over the AG equation, since the solutions of \eqref{UpsEq} are defined on a lower-dimensional space than the solutions of the AG equation.}

\section{Discussion}
In this section, we discuss some of the consequences and implications of the classical-quantum wave equation \eqref{UpsEq}. 
The first observation is about quantum decoherence, which naturally arises from the first in \eqref{PartialTraces} in terms of purity non-preservation. Also, we observe that classical dynamics can be different from what we are used to in the absence of classical-quantum interaction. On one hand, the last equation in \eqref{PartialTraces} does not generally allow for point particle solutions. Since the latter are known to be classical pure states \cite{ChernoffMarsden,Shirokov}, we conclude that classical-quantum correlations induce a loss of classical purity that mimics quantum decoherence effects. This will be illustrated below on an exactly solvable example. 

On the other hand, as we pointed out, positivity of $\rho$ may not be generally preserved in time \cite{boucher}. Indeed, while the sign of $\rho$ will be shown to be preserved for certain classes of hybrid Hamiltonians $\widehat{H}$ (see Section \ref{sec:example}), it is not possible to draw a similar conclusion in the general case.
Although the emergence of a sign-indefinite $\rho$ may seem surprising at first, an analogue of this situation can be readily found in the standard case of a harmonic oscillator interacting (by a linear or quadratic coupling) with a nonlinear quantum system. Let us consider the full quantum case in the Wigner representation: the Wigner-Moyal equation for $W(\bz,\bzeta)$ reads
\[
{\partial_t} W = \{\!\!\{H,W\}\!\!\}_\bz + \{\!\!\{H,W\}\!\!\}_\bzeta,
\] 
where $\{\!\!\{\,,\,\}\!\!\}$ denotes the Moyal bracket in the set of coordinates given by the subscript. Here, $H(\bz,\bzeta)$ retains arbitrary nonlinear dependence on $\bzeta$, while it is quadratic in $\bz$ so that $\{\!\!\{H,W\}\!\!\}_\bz=\{H,W\}_\bz$. We emphasize that, in the absence of the nonlinear quantum system, we have $\nabla_\bzeta H=0$ and the oscillator undergoes  classical evolution (while its quantum features are encoded in the initial condition). This means that the coupled  system can be considered as equivalent to a hybrid classical-quantum system. 
Then, projecting out the quantum coordinates yields an equation for $\varrho(\bz) =  \int \!W(\bz,\bzeta)\, \de^6\zeta$, that is ${\partial_t}\varrho  = \int \{H,W\} \,\de^6\zeta$.
This is exactly the analogue of our second equation in \eqref{PartialTraces}. Also in this case, despite the classical structure of the oscillator subsystem, its density $\varrho$ may develop negative values in time (even if $\varrho > 0$ initially) because $W$ is not generally positive. Then, as already pointed out by Feynman \cite{FeynmanNegativeProb}, the possibility  of nonpositive classical distributions in compound systems does not come as a surprise. Further discussions on the meaning of negative probabilities and their applications can be found, e.g., in \cite{FeynmanNegativeProb, Khrennikov}.

In addition, we wish to emphasize that, unlike Sudarshan's model \cite{Sudarshan}, the present construction consistently recovers the mean-field model for the classical and quantum densities. This is readily verified by replacing the mean-field factorization ansatz $\Upsilon(\bz,\boldsymbol{x})=\Psi(\bz)\psi(\boldsymbol{x})$ in the variational principle \eqref{hybridVP}. Indeed, this operation returns
\begin{align}\label{MF1}
i\hbar\partial_{t\!}\Psi=&\, 
%\widehat{\cal L}_{\langle \psi|\widehat{H}\psi\rangle}\Psi=
\langle\psi|\widehat{H}\psi\rangle\Psi-\nabla \langle\psi|\widehat{H}\psi\rangle\cdot\,\widehat{\!\boldsymbol{\cal Z}\,}_{\!\!+}\Psi
\\
i\hbar\partial_t\psi=&\,\bigg(\int\!\Psi^*\widehat{\cal L}_{\widehat{H}}\Psi\,\de^6z\bigg)\psi
\,,
\label{MF2}
\end{align}
so that the equations for the quantum density $\hat\rho=\psi\psi^\dagger$ and the classical distribution $\rho$ (as given in \eqref{Clebsch}) return the mean-field equations in the form
\beq
{\partial_t}\rho=\{\operatorname{\sf Tr} (\hat{\rho}\widehat{H}),\rho \}
\,,\qquad \quad
i\hbar{\partial_t\hat\rho}=\bigg[\int\!\rho\widehat{H}\,\de^6z\,,\hat{\rho}\bigg]
.
\label{MFeqs}
\eeq
{We emphasize that here the mean-field model emerges as an exact closure obtained from the variational structure \eqref{hybridVP} of the classical-quantum wave equation \eqref{UpsEq}. The same does not hold for the AG equation \eqref{Aleq}, which indeed lacks a variational formulation. As shown in \cite{Gerasimenko}, replacing the mean-field factorization ansatz $\widehat{\cal D}(\bz,t)=\hat\rho(t)\,\rho(\bz,t)$ in \eqref{Aleq} yields an unclosed system, which then requires the extra closure condition of vanishing classical-quantum correlations. 
}

Before concluding this section, it may be relevant to highlight that the whole construction presented here can also be reformulated in terms of a density-like operator. Indeed, one can simply replace the classical-quantum wave equation \eqref{UpsEq} by its correspondent for a positive-definite density-like operator $\widehat{\Theta}$, that is
\[
i\hbar{\partial_t\widehat{\Theta}}=\big[\widehat{\cal L}_{\widehat{H}},\widehat{\Theta}\big]
\,,
\]
which we shall call \emph{classical-quantum von Neumann equation}.
Given the level of difficulty of such an extension of the theory, in this paper we choose to leave this direction open for future work.

%%%%%%%%%%%%%%%%%%%%%%%%

\section{An exactly solvable system\label{sec:example}}

Many studies utilize a linear classical-quantum interaction potential preventing quantum backreaction beyond mean-field effects. Indeed, in these cases the force exerted on the classical degrees of freedom by the quantum subsystem does not depend on classical-quantum correlations. For example, in the case of the Jaynes-Cummings model, the expectation value dynamics for the classical momentum depends only on the spin expectation $\langle \widehat{\boldsymbol\sigma}\rangle$ (already occurring in the mean-field model), but not on mixed quantum-classical expectations, e.g., $\langle q\widehat{\boldsymbol\sigma}\rangle$. For the latter term to appear in the equation of the classical momentum expectation, a quadratic coupling between the classical and quantum subsystems is needed. Hence, to demonstrate the emergence of the quantum backreaction, we consider the exactly solvable case of a degenerate two-level quantum system quadratically coupled to a one-dimensional classical harmonic oscillator. The Hamiltonian of such a system reads 
\beq\label{EqHamiltonianExactSolvableSystem}
	\widehat{H} = H_{0} +\frac{q^2}{2} \boldsymbol\alpha\cdot\widehat{\boldsymbol\sigma}, \qquad\   H_{0} = \frac{p^2 }{ 2m} + m\omega^2 \frac{q^2}{2}.
\eeq
Here, $m$ and $\omega$ denote respectively the mass and frequency of the harmonic oscillator, $\widehat{\sigma}_j$ are the Pauli matrices ($j=1,2,3$) representing the two-level quantum system, and the  vector $\boldsymbol\alpha$ comprises the classical-quantum coupling constants  $\alpha_j$. {Since this example involves a harmonic oscillator, here we shall adopt the convenient gauge \eqref{HOg}. In this case,} the hybrid equation of motion \eqref{UpsEq} reads
\beq\label{EqHarmonicOscBeforeRotation}
	\frac{\partial\Upsilon}{\partial t}  = \left[ q\left(m\omega^2 +  \boldsymbol\alpha\cdot\widehat{\boldsymbol\sigma} \right) \frac{\partial}{\partial p}  - \frac{p}{m} \frac{\partial}{\partial q}  \right] \Upsilon,
\eeq
where $\Upsilon =(\Upsilon_1(q,p,t),\Upsilon_2(q,p,t))^T\in\mathbb{C}^2$. The equations for each component are decoupled after introducing the wavefunction $\widetilde{\Upsilon}=\widehat{U}\Upsilon$, where the unitary matrix $\widehat{U}$ is defined by { $\widehat{U} {(\boldsymbol\alpha\cdot\widehat{\boldsymbol\sigma})} \widehat{U}^{\dagger} = \lambda \widehat{\sigma}_3$. In the last equation, we have used the fact that the matrix ${\boldsymbol\alpha\cdot\widehat{\boldsymbol\sigma}}$ is traceless thus its eigenvalues must be of equal magnitude but with the opposite sign.} Then, solving each linear characteristic equation for each component $\widetilde{\Upsilon}_k$ leads to the following exact solution of \eqref{EqHarmonicOscBeforeRotation}, expressed in terms of the initial condition $\Upsilon_0 = \Upsilon|_{t=0}$:
{
\begin{align}\label{EqExactSolutionHybrid}
	\Upsilon = \widehat{U}^{\dagger} \left( y_1(\omega_{+}) \atop y_2(\omega_{-}) \right),
\end{align}
where $\omega_{\pm} = \sqrt{\omega^2 \pm\lambda / m}$ and $y_l(\omega_{\pm})$ denotes the component of the vector
\begin{align}
	y(\omega_{\pm}) = \widehat{U} \Upsilon_0 \left(  
					q=q\cos(\omega_{\pm} t) - \frac{p \sin(\omega_{\pm} t)}{m\omega_{\pm}},
					p=p\cos(\omega_{\pm} t) + m \omega_{\pm} q \sin(\omega_{\pm} t)
				\right).
\end{align}
}

\begin{figure}
\center
	\includegraphics[width=.55\hsize]{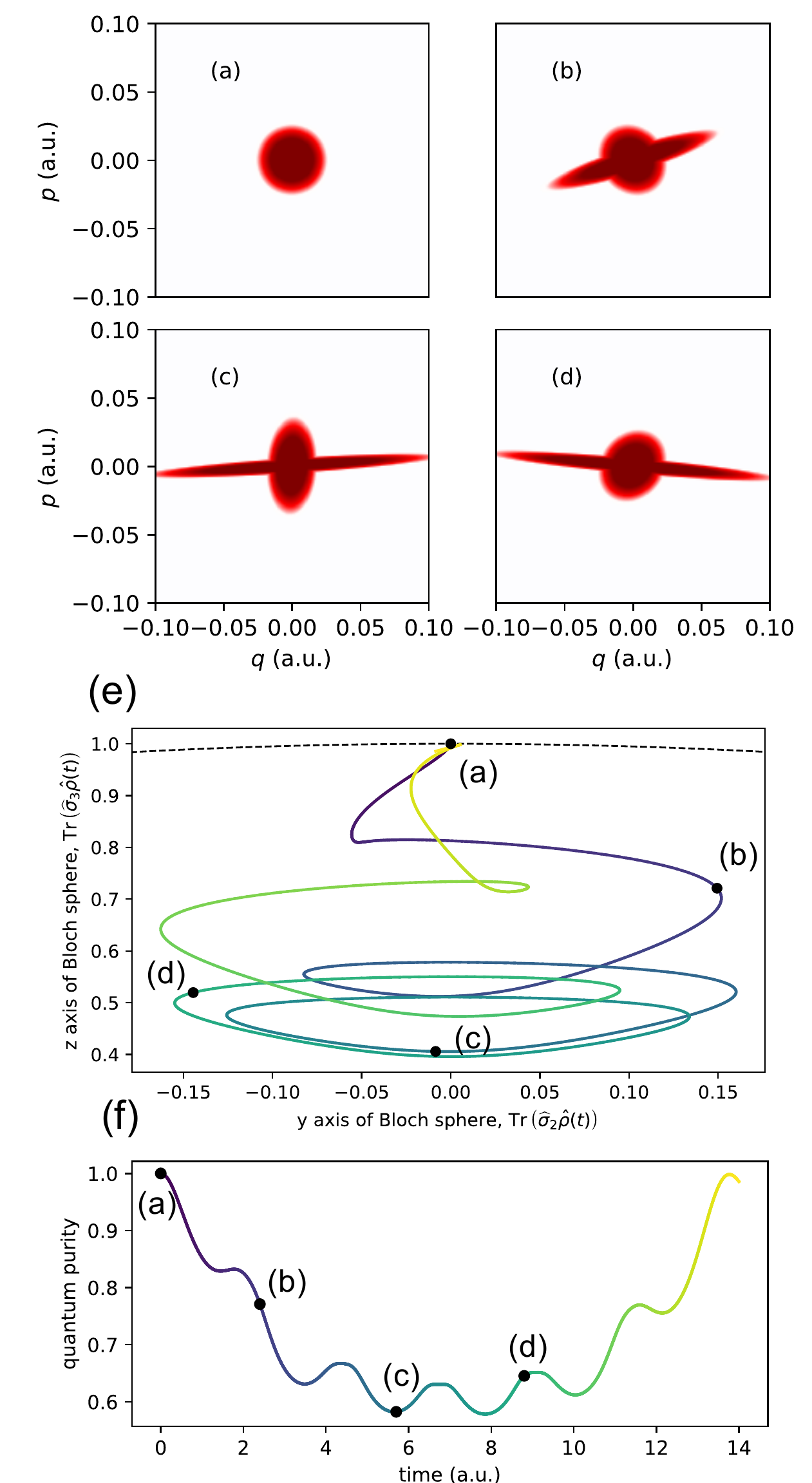}
	\caption{\footnotesize Hybrid evolution of a degenerate two-level quantum system quadratically coupled to a one-dimensional classical harmonic oscillator. The system Hamiltonian is given in \eqref{EqHamiltonianExactSolvableSystem}. The depicted dynamics has the the exact solution \eqref{EqExactSolutionHybrid}  with $\omega = m = 1$ (a.u), $\boldsymbol\alpha = (0.95,0,0)$ (a.u.), and the factorized initial condition \eqref{EqInitialCondition} with $\beta = 10^{5}$ (a.u.). The classical Liouville density \eqref{EqClassicalDensity} for this system is depicted at different times $t = 0, \, 2.4, \, 5.7, \, 8.8$ (a.u.) in figures (a), (b), (c), and (d), respectively. Red colors corresponds to positive values of the classical density $\operatorname{\sf Tr} \widehat{\cal D}$, whereas white marks vanishingly small values. Figure (e) depicts the trajectory traced by the Bloch vector $\boldsymbol{n}=\operatorname{\sf Tr}(\widehat{\boldsymbol\sigma}{\hat\rho})$ for the quantum density matrix \eqref{EqQuantumDensity} during the evolution. The progression of time is represented by a color gradation from dark blue to yellow along the curve. Since the trajectory lies on the $yz$ plane, only the $yz$ projection is plotted. The dashed back line denotes the surface of the Bloch sphere. Figure (f) displays the purity $\operatorname{\sf Tr}({\hat\rho}^2)=|\boldsymbol{n}|^2$ of the quantum density matrix \eqref{EqQuantumDensity} as a function of time. In figures (e) and (f) the captioned black dots mark time at which figures (a)-(d) are plotted. The color encoding of time is the same in both figures (e) and (f).}\label{FigExactSolution}
\end{figure}

Figure \ref{FigExactSolution} depicts the classical-quantum evolution for such a system with the initial condition
\beq\label{EqInitialCondition}
	\Upsilon_0 = \sqrt{\frac{\omega}{2\pi}\frac{1 - \left(1 + \beta H_{0}\right) e^{-\beta H_{0}}}{\beta H_{0}^2} }\, \left( 1 \atop 0 \right), \qquad\qquad
	\widehat{\cal D}_0 =  \frac{\omega\beta}{2\pi} e^{-\beta H_0} \left( 1 \ \  0 \atop 0  \ \ 0 \right),
\eeq
corresponding to the uncorrelated quantum-classical state, where the quantum state \eqref{EqQuantumDensity} is the ground (i.e., ``up'') state and the classical Liouville density \eqref{EqClassicalDensity} is the Boltzmann state $\rho \propto e^{-\beta H_{0}}$  with an inverse thermodynamic temperature $\beta$ and $H_{0}$ as given in \eqref{EqHamiltonianExactSolvableSystem}. The long-tailed  wavefunction $\Psi$ given by the square root in \eqref{EqInitialCondition} and corresponding to the classical Boltzmann state can be easily obtained upon recalling \eqref{Clebsch} and by solving the differential equation $|\Psi|^2+\operatorname{\sf div}\!\big(\Psi^*\widehat{\!\boldsymbol{\cal Z}\,}_{\!\!+}\Psi\big)=\omega\beta e^{-\beta H_{0}}/2\pi$. The latter is taken into a linear first-order ODE for $|\Psi|^2$ by setting a zero phase and then changing to polar coordinates. We remark that the initial condition \eqref{EqInitialCondition} represents a stationary state for the uncoupled classical-quantum system, that is $\boldsymbol\alpha = 0$. {See Proposition 22.6 in \cite{Hall} for the characterization of the stationary states of the KvH equation for the harmonic oscillator}.

 Figure \ref{FigExactSolution} uses the atomic units (a.u.), where the electron mass, the electron charge, and $\hbar$ are all set to a unity \cite{Material}.   As it can be seen, the quantum-classical correlations rapidly develop yielding non-Gaussian classical Liouville densities \eqref{EqClassicalDensity} (due to the quantum backreaction) and non-pure quantum states \eqref{EqQuantumDensity}. In other words, the classical system induces quantum decoherence [see figure \ref{FigExactSolution}(f)]. It is noteworthy that the classical density is non-negative for all times in the considered example{; we shall expand this particular point at the end of this section.}

\begin{figure}[h]
\center
	\includegraphics[width=.52\hsize]{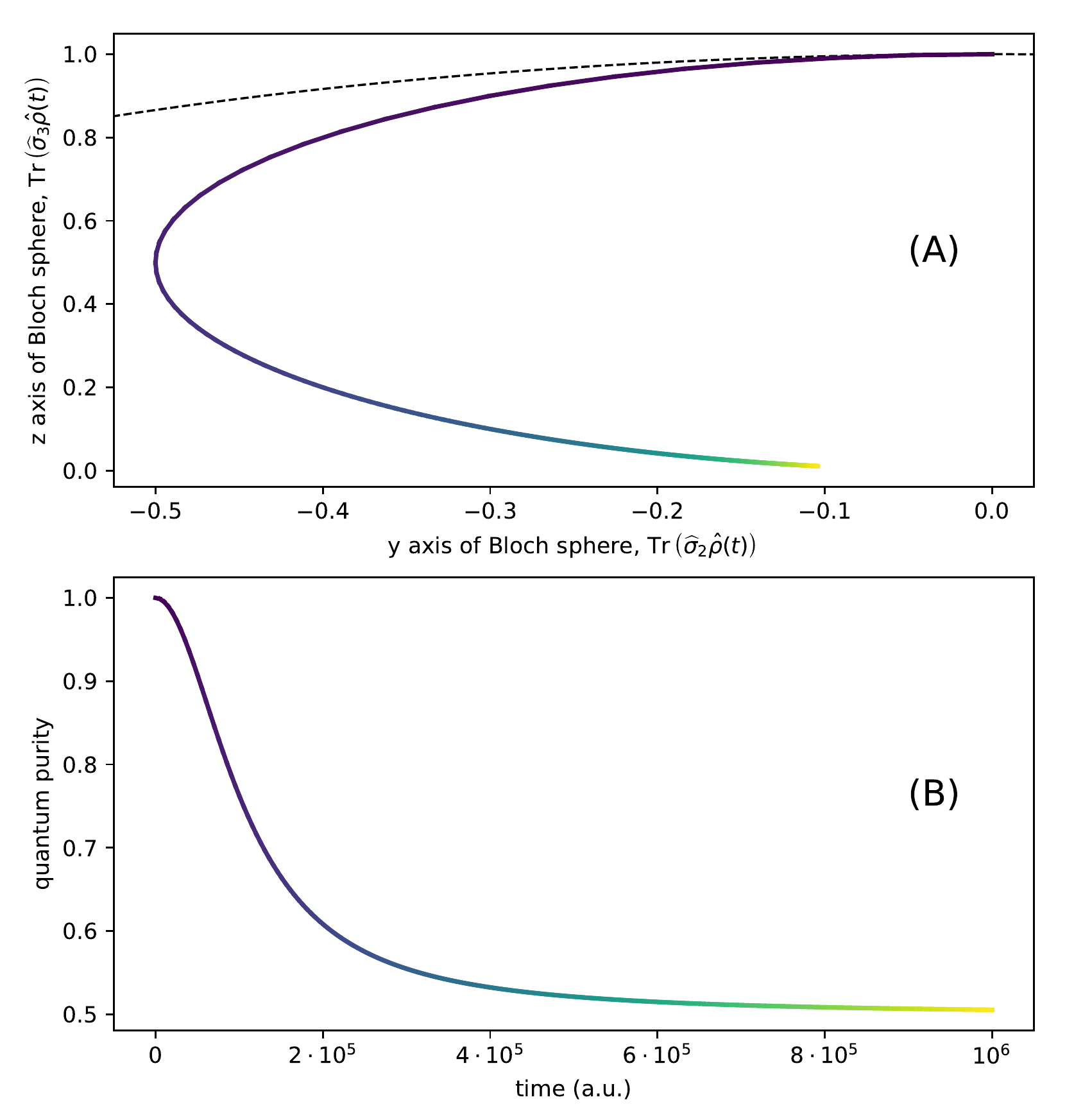}
	\caption{\footnotesize Hybrid evolution \eqref{EqExactSolAG} governed by the AG equation \eqref{Aleq} with the Hamiltonian given in \eqref{EqHamiltonianExactSolvableSystem} and the initial condition $\widehat{\cal D}_0$ in \eqref{EqInitialCondition}. The parameters used are the same as in figure \ref{FigExactSolution}. Figure (A) depicts the trajectory traced by the Bloch vector for the quantum density matrix $\hat{\rho} = \int\! \widehat{\cal D} dpdq$ during the evolution. Similarly to figure  \ref{FigExactSolution}, the progression of time is represented by a color gradation from dark blue to yellow. Again, the trajectory lies on the $yz$ plane. However, we emphasize the very different time scale from the evolution displayed in figure \ref{FigExactSolution}.   Figure (B) displays the purity $\operatorname{\sf Tr}({\hat\rho}^2)$ as a function of time. The color encoding of time is the same in both figures (A) and (B). The classical Liouville density $\operatorname{\sf Tr} \widehat{\cal D}$ is identical to the top four panels in \makebox{figure \ref{FigExactSolution}.}}\label{Fig2}
\end{figure}

{
It is instructive to compare these findings with the predictions of the AG theory \eqref{Aleq}. The exact solution of the AG equation \eqref{Aleq} for the Hamiltonian \eqref{EqHamiltonianExactSolvableSystem} reads in terms of the initial condition $\widehat{\cal D}_0 = \widehat{\cal D}|_{t=0}$ as
\begin{align}\label{EqExactSolAG}
	\widehat{\cal D} = \widehat{U}^{\dagger} \left(
		\begin{array}{cc}
			d_{11}(\omega_{+}) & e^{i\varphi} d_{12}(\omega) \\
			e^{-i\varphi} d_{21}(\omega) & d_{22}(\omega_{-}) 
		\end{array}
	\right) \widehat{U},
\end{align}
where $d_{kl}(\omega_{\pm})$ denote the components of the matrix 
\beq
	\widehat{d}(\omega_{\pm}) = \widehat{U} \widehat{\cal D}_0 \left(
		q=q\cos(\omega_{\pm} t) - \frac{p \sin(\omega_{\pm} t)}{m\omega_{\pm}},
		p=p\cos(\omega_{\pm} t) + m \omega_{\pm} q \sin(\omega_{\pm} t)
	\right)\widehat{U}^{\dagger},
\eeq
and
\beq\label{EqPhiAG}
	\varphi = \frac{\lambda}{2m\hbar \omega^3}\left(
		\frac{p^2 - (m\omega q)^2}{2m} \sin(2\omega t)	 
		-\omega \left[ 2H_0t + pq \left( \cos(2\omega t) - 1 \right) \right]
	\right).
\eeq

The exact solutions \eqref{EqExactSolutionHybrid} and \eqref{EqExactSolAG} lead to qualitatively different dynamics. In particular, the phase $\varphi$ breaks the time-reversible symmetry in AG hybrid dynamics. Furthermore, the term $2H_0t$ in \eqref{EqPhiAG} yields a non-periodic evolution, which is responsible for the purity relaxation at large timescales, as  shown in figure \ref{Fig2}. 
The density matrix of the quantum subsystem monotonically approaches an infinite-temperature state. This dynamics is reminiscent of the relaxation predicted by the Lindblad  equation modeling a dephasing channel. Indeed, in the case of the Lindblad  equation, the entropy-driven relaxation process at macroscopic time-scales
(such as those in figure \ref{Fig2}) is predicted by an H-theorem \cite{Lindblad2}.
However, the lack of any features at microscopic timescales prevents the AG equation from capturing transient behavior. This should be contrasted with the predictions of the new model depicted in figure \ref{FigExactSolution}, where recurrent quasi-periodic dynamics, akin to the Rabi oscillations, is observed with no long-time trend -- a direct consequence of the model having the Hamiltonian structure. 
Despite these substantial differences, we emphasize that both the solutions \eqref{EqExactSolutionHybrid} and \eqref{EqExactSolAG} lead to the same classical Liouville density as shown in the top four panels of figure \ref{FigExactSolution}. Another similarity between the two theories is that they both produce   negative eigenvalues of the hybrid density $\widehat{\cal D}$. This fact was numerically verified for the considered example.

The parameters chosen in figures \ref{FigExactSolution} and \ref{Fig2} are such that  $\beta \gg 2 / (\hbar\omega)$. This means that the initial condition \eqref{EqInitialCondition}  identifies a cold classical state, whose phase space distribution  in figure \ref{FigExactSolution}(a) violates the Heisenberg uncertainty principle. Therefore, figure \ref{FigExactSolution} and \ref{Fig2} display truly hybrid dynamics, rendering quantum-classical correlations that cannot be modeled by the Pauli equation. However, if we set $\beta = 2 / (\hbar\omega)$, the initial classical Liouville density  $\rho$ coincides with the Wigner function $(\pi\hbar)^{-1\!}\int\psi^\dagger(q+s)\psi(q-s)\,e^{2ips/\hbar}\,\de s$ for the Pauli spinor wavefunction $\psi(q) \propto e^{-m\omega q^2/(2\hbar)} (1, 0)^T$. For such an initial condition, the classical density dynamics arising from equations of motion \eqref{Aleq} and \eqref{EqHarmonicOscBeforeRotation} coincide with the evolution of the Wigner function associated to the Pauli equation with Hamiltonian $\hat{p}^2 / (2m) + m\omega^2 \hat{q}^2/2 + \boldsymbol\alpha\cdot\widehat{\boldsymbol\sigma} \hat{q}^2/2$, where $[\hat{q}, \hat{p}] = i\hbar$. 

We conclude this section by showing that any hybrid Hamiltonian of the type $\widehat{H}(q,p)  = H_{0}(q,p)  +V(q) \boldsymbol\alpha\cdot\widehat{\boldsymbol\sigma}$ yields a hybrid wave equation \eqref{UpsEq} that preserves the sign of the classical Liouville density. By following the diagonalization procedure above, this class of hybrid Hamiltonians can be equivalently written as $\widehat{H} = H_{0} +\lambda \widehat{\sigma}_3 V$, thereby producing two uncoupled KvH equations $i\hbar{\partial_t\widetilde{\Upsilon}_{\pm}}  =  \widehat{\mathcal{L}}_{H_{\pm}}\widetilde{\Upsilon}_{\pm}$ of classical type (here, ${H_{\pm}}=H_0 \pm \lambda V$). From the arguments in Section \ref{Sec:KvH}, it follows that both these KvH equations preserve the sign of the quantity $\rho_{\pm}=|\widetilde{\Upsilon}_{\pm}|^2+\operatorname{\sf div}(\widetilde{\Upsilon}^*_{\pm\ }\widehat{\!\boldsymbol{\cal Z}\,}_{\!\!+}\widetilde{\Upsilon}_{\pm})$. As a result, the sign of the classical density  $\rho=\rho_{+} +\rho_{-}$ of the hybrid system is also preserved in time. This result promises well for possible other classes of hybrid Hamiltonians yielding positivity of the classical distribution; such a study is the subject of ongoing work \cite{GBTr19}.
}

\section{Conclusions}
Upon combining KvN  classical mechanics with van Hove's prequantization theory, we have provided the new representation \eqref{Clebsch} of the Liouville density in terms of Koopman-van Hove classical wavefunctions. Then, given the KvH equation \eqref{modKvN} for two particles, a quantization procedure was applied to one of them thereby leading to the classical-quantum wave equation \eqref{UpsEq} for the hybrid wavefunction $\Upsilon(\bz,\bx)$. This construction leads naturally to the identification of  a sign-indefinite operator-valued density \eqref{Ddef} encoding classical-quantum correlations. In turn, the latter can be discarded by invoking the factorization ansatz $\Upsilon(\bz,\bx)=\Psi(\bz)\psi(\bx)$ recovering the celebrated mean-field model \eqref{MFeqs}.

 Equations \eqref{UpsEq}, \eqref{Ddef}, \eqref{EqQuantumDensity}, and \eqref{EqClassicalDensity} constitute a long sought Hamiltonian model for classical-quantum hybrid evolution. As shown, the density matrix of the quantum subsystem is always positive, while the Liouville density of the classical subsystem may, in general, become negative in the general case.   The proposed hybrid description has been illustrated { and compared to the AG theory \eqref{Aleq}} by using the exactly solvable model of a degenerate two-level quantum system quadratically coupled to a one-dimensional classical harmonic oscillator. In this case, the quantum backreaction leads to positive-definite, yet non-Gaussian classical distributions. The discussion of which classes of hybrid systems preserve the sign of the classical distribution is left for future work \cite{GBTr19}. Other questions currently under study \cite{GBTr19} involve the algebraic structure of the hybrid correspondence ${\widehat{H}}\to\widehat{\cal L}_{\widehat{H}}$ and the associated dual map yielding the hybrid density $\widehat{\cal D}$.
 
As a further  direction, we plan to develop effective numerical schemes for the   classical-quantum wave equation \eqref{UpsEq}  to be able to assess its physical consequences in experimentally relevant scenarios, such as those involving the Jaynes-Cummings model. In addition,  the identification of hybrid classical-quantum thermal equilibria is an interesting question whose answer may open new perspectives in the statistical mechanics of hybrid classical quantum systems \cite{Ciccotti}.
Indeed, once a Hamiltonian model is established, the immediate next question involves its extension to time-irreversible processes governed by an H-theorem. We remark that time irreversibility and energy dissipation are substantially different phenomena which may or may not coexist. Examples are given by the quantum Lindblad equation and the classical Botzmann equation, respectively. The addition of thermodynamic effects to Hamiltonian theories is a challenging question requiring  methods from statistical mechanics.   We leave this important direction for future work.

\medskip

\paragraph{Acknowledgments.} {We thank all six referees for their careful reading of the manuscript and for their keen remarks that contributed to improving the exposition of our results}. We are grateful to Darryl D. Holm for his valuable comments during the writing of this work. Special thanks go to Paul Skerritt for his keen insight into the identification of specific initial conditions for the classical wavefunction. Also, the authors are  indebted to Dorje Brody, Joshua Burby, Maurice de Gosson, Hans-Thomas Elze, Viktor Gerasimenko, Raymond Kapral, Robert Littlejohn, Robert MacKay, Omar Maj, Giuseppe Marmo, Todd Martinez, Philip Morrison, Jonathan Oppenheim, Hong Qin, Milan Radonji\'c, Lorenzo Salcedo, and Ivano Tavernelli for stimulating correspondence and inspiring discussions. D.I.B. is supported by Air Force Office of Scientific Research Young Investigator Research Program (No. FA9550-16-1-0254). F.G.B was partially supported by the grant ANR-14-CE23-0002-01.  
C.T. acknowledges financial support from the Leverhulme Trust Research Project Grant No. 2014-112, and from the London Mathematical Society Grant No. 31633 (Applied Geometric Mechanics Network). This material is partially based upon work supported by the NSF Grant No. DMS-1440140 while C.T. was in residence at MSRI, during the Fall 2018 semester. In addition, both D.I.B. and C.T. acknowledge support from the Alexander von Humboldt Foundation (Humboldt Research Fellowship for Experienced Researchers) as well as from the German Federal Ministry for Education and Research.

\appendix

\section{KvH momentum map}\label{Appendix_A}

In this Section, we show explicitly that the relation \eqref{Clebsch} identifies a momentum map for the infinitesimal action given by the operator $\widehat{\cal L}_H$. In Geometric Mechanics \cite{MaRa,HoScSt}, momentum maps \cite{Sternberg,Sternberg2} represent a generalization of Noether's theorem to canonical group actions that are not necessarily a symmetry of the system under consideration. In this context, the Noether charge is generalized to a momentum map that evolves under the coadjoint representation associated to the  Lie group acting on the considered mechanical system. 

Without entering further details, we define the momentum map on symplectic vector spaces as follows. Let $(V,\Omega)$ be a vector space with {constant} symplectic form $\Omega$ and let the latter be preserved by a $G-$group {representation} on $V$. Then, the momentum map ${\bf J}:V\mapsto\mathfrak{g}^*$ taking values in the dual space $\mathfrak{g}^*$ of the Lie algebra $\mathfrak{g}$ of $G$ is defined as
\[
2\langle{\bf J}(v),\xi\rangle:=\Omega(\xi_V(v),v)
\,,
\]
where $\xi\in \mathfrak{g}$, $\xi_V$ denotes the infinitesimal action on $V$, and $\langle\cdot,\cdot\rangle$ is the real-valued duality pairing for $ \mathfrak{g}$. The momentum map ${\bf J}(v)$ is generally called a \emph{Clebsch representation}.

 In our case,  $V$ is the space of classical wavefunctions, the Lie algebra is the space $\mathfrak{g}=C^\infty(\Bbb{R}^6)$ of phase-space functions (endowed with the canonical bracket and the standard $L^2-$pairing), and the infinitesimal generator $\xi_V(v)$ reads $-i\hbar^{-1}\widehat{\cal L}_H\Psi$. Then, upon using the Schr\"odinger (canonical) symplectic form $\Omega(\Psi_1,\Psi_2)=2\hbar\operatorname{\sf Im}\int\!\Psi_1^*(\bz)\Psi_2(\bz)\,\de^6z$, the definition of momentum map reads
\[
\int\! H{\bf J}(\Psi)\,\de^6z
=
\int\!\Psi^*\widehat{\cal L}_H\Psi\,\de^6z\,.
\]
Therefore, we compute
\begin{align*}
\int\!\Psi^*\widehat{\cal L}_H\Psi\,\de^6z=&
\int\!\Psi^*\Big[\{i\hbar H,\Psi\}
+\Big(H-\boldsymbol{\cal A}\cdot J\nabla H\Big)\Psi\Big]\,\de^6z
\\
=&
\int\!
\Big[|\Psi|^2-\operatorname{\sf div}\big(J\boldsymbol{\cal A}|\Psi|^2\big)
+
i\hbar\{\Psi,\Psi^*\}\Big]H\,\de^6z\,.
\end{align*}
Now, we observe that 
$
i\hbar\{\Psi,\Psi^*\}=-i\hbar\operatorname{\sf div}(\Psi^*J\nabla \Psi)=\operatorname{\sf div}(\Psi^*J\widehat{\bLambda} \Psi)
$
so that the momentum map reads
\[
{\bf J}(\Psi)=|\Psi|^2+\operatorname{\sf div}\!\big[\Psi^*J\big(\widehat{\bLambda}-\boldsymbol{\cal A}\big)\Psi\big]
,
\]
thereby recovering the relation \eqref{Clebsch}  as a Clebsch representation. By proceeding analogously, we notice that $|\Psi|^2$ is also a Clebsch representation generated by local phase transformations with infinitesimal action $\xi_V(v)$ given as $-i\hbar^{-1}\phi\Psi$ (where $\phi(\bz)$ is a real phase-space function).

Notice that, since $-i\hbar^{-1}\widehat{\cal L}_H$ is skew-Hermitian, the correspondence $H\mapsto-i\hbar^{-1}\widehat{\cal L}_H$ provides a Lie algebra homomorphism between phase-space functions {endowed with the canonical Poisson bracket} and skew-Hermitian operators on classical wavefunctions. Then, the map  $-i\hbar\Psi(\bz)\Psi^*(\bz')\mapsto {\bf J}(\Psi)$ emerges as the dual of this Lie algebra homomorphism, thereby ensuring infinitesimal equivariance of $\mathbf{J}(\Psi)$ and the consequent Poisson mapping property \cite{MaRa,HoScSt}. Thus, this guarantees that the momentum map ${\bf J}(\Psi)$ obeys the classical Liouville equation. Again, without entering further details, here we only mention that the operator  $-i\hbar^{-1}\widehat{\cal L}_H$ emerges as the infinitesimal generator of a Lie group {representation} first discussed in Van Hove's thesis \cite{VanHove}, which is at the heart of classical mechanics. Under the name of ``strict contact transformations'', this Lie group is a central extension of standard canonical transformations. This and related points will be discussed in more detail in future work.
%endowed with the $L^2$ pairing $\langle\Psi_1,\Psi_2\rangle=\operatorname{\sf Re}\langle\Psi_1|\Psi_2\rangle=\operatorname{\sf Re}\!\int\!\Psi_1^*(\bz)\Psi_2(\bz)\,\de^6z$

{
An important consequence of the fact that the operator $-i\hbar^{-1}\widehat{\cal L}_H$ generates strict contact transformations (as opposed to $-i\hbar^{-1}\widehat{L}_H$, which generates canonical transformations) is the  
 equivariance property resulting as a general property of infinitesimal generators associated to group actions. As discussed in Section \ref{Sec:KvH}, in the Heisenberg picture 
  we have $\widehat{\cal L}_A^{\,\sf H}= \widehat{\cal L}_{A^{\sf H}}$ where ${\widehat{\cal L}}_A^{\,\sf H}(t):=\exp({i{\widehat{\cal L}}_H}t/\hbar){\widehat{\cal L}}_A \exp({-i{\widehat{\cal L}}_H}t/\hbar)$ and $A^{\sf H}(t)=\exp({i{\widehat{L}}_H}t/\hbar)A$. Indeed,  this is a consequence of the general formula $(\operatorname{Ad}_{g}\xi)_V= \Phi_{g^{-1}}^*\xi_V$ for a left representation $\Phi$ of a Lie group $G$ on a vector space $V$. In the specific case under consideration, the adjoint action $\operatorname{Ad}_{g}\xi$ coincides with the pushforward $A\circ\eta^{-1}$ of the function $\xi=A$ by the canonical transformation $\eta^{-1}=\exp(-{X}_H t)$ generated by $i\hbar^{-1}\widehat{L}_H$, so that $\operatorname{Ad}_{\eta}A=A\circ\eta^{-1}$. Therefore, we write $A\circ\eta^{-1}=\exp({i{\widehat{L}}_H}t/\hbar)A$.}

\section{Hybrid  dynamics}

In this Appendix, we provide calculational details of the discussion concerning classical-quantum hybrids. {Here $\boldsymbol{\cal A}$ is an arbitrary potential with ${\rm d}\boldsymbol{\cal A}=-\omega$ or, equivalently, $\nabla\boldsymbol{\cal A}-(\nabla\boldsymbol{\cal A})^T=-J$.}
First, we shall show that the definition \eqref{Ddef} leads to rewriting the total energy  \eqref{TotEn} as
\[
h=\operatorname{\sf Tr}\!\int\!\Upsilon^\dagger(\bz)\,\widehat{\cal L}_{\widehat{H}}\Upsilon(\bz)\,\de^6z
=\operatorname{\sf Tr}\!\int\!\widehat{H}\widehat{\cal D}\,\de^6z
\,,
\]
with ${\cal D}$ given in  \eqref{Ddef}.
Indeed, we verify this as follows:
\begin{align*}
\operatorname{\sf Tr}\!\int\!\Upsilon^\dagger\,\Big[\widehat{H}-
\nabla\widehat{H}\cdot J\big(\widehat{\bLambda}-\boldsymbol{\cal A}\big)\Big]\Upsilon\,\de^6z
&\ =
\operatorname{\sf Tr}\!\int\!\Big[\Upsilon\Upsilon^\dagger\widehat{H}-
\operatorname{\sf div}(J\boldsymbol{\cal A}\Upsilon\Upsilon^\dagger)\widehat{H}
+i\hbar
\Upsilon^\dagger\{\widehat{H},\Upsilon\}\Big]\de^6z
%\\
%&\ =
%\operatorname{\sf Tr}\!\int\!\Big[\Upsilon\Upsilon^\dagger\widehat{H}+
%\operatorname{\sf div}\Big(\frac\bz2\Upsilon\Upsilon^\dagger\Big)\widehat{H}\Big]\,\de^6z
%\\
%&\qquad +i\hbar
%\operatorname{\sf Tr}\!\int\!\widehat{H}\{\Upsilon,\Upsilon^\dagger\}\,\de^6z
%\\
%&\ =
%\operatorname{\sf Tr}\!\int\!\Big[\Upsilon\Upsilon^\dagger\widehat{H}+
%\operatorname{\sf div}\Big(\frac\bz2\Upsilon\Upsilon^\dagger\Big)\widehat{H}\Big]\,\de^6z
%\\
%&\qquad +
%\operatorname{\sf Tr}\!\int\!\widehat{H}\operatorname{\sf div}(\Upsilon(i\hbar J\nabla)\Upsilon^\dagger)\,\de^6z
\\
&\ =
\operatorname{\sf Tr}\!\int\!\Big[\Upsilon\Upsilon^\dagger\widehat{H}-
\operatorname{\sf div}(J\boldsymbol{\cal A}\Upsilon\Upsilon^\dagger)\widehat{H}
-
\widehat{H}\operatorname{\sf div}(\Upsilon J\widehat{\bLambda}\Upsilon^\dagger)\Big]\de^6z
\\
&\ =
\operatorname{\sf Tr}\!\int\!\Big[\Upsilon\Upsilon^\dagger+
\operatorname{\sf div}(\Upsilon\,\widehat{\!\boldsymbol{\cal Z}\,}_{\!\!-}\Upsilon^\dagger)\Big]\widehat{H}\,\de^6z
\,,
\end{align*}
where all quantities are evaluated at $\bz$ and we  used
\begin{align*}
i\hbar\!
\operatorname{\sf Tr}\!\int\!\Upsilon^\dagger\{\widehat{H},\Upsilon\}\,\de^6z
=&\, 
i\hbar\!
\operatorname{\sf Tr}\!\int\!\widehat{H}\{\Upsilon,\Upsilon^\dagger\}\,\de^6z
=
\operatorname{\sf Tr}\!\int\!\widehat{H}\operatorname{\sf div}(\Upsilon(i\hbar J\nabla)\Upsilon^\dagger)\,\de^6z
\,.
\end{align*}
In conclusion, we recover the definition  \eqref{Ddef}.

Now we want to prove the $\widehat{\cal D}-$equation \eqref{Deqn}. For this purpose, we shall use the adjoint of equation \eqref{UpsEq}, that is 
\[
-i\hbar\partial_t\Upsilon^\dagger(\bz)=\Upsilon^\dagger(\bz)\widehat{H}(\bz)-(\,\widehat{\!\boldsymbol{\cal Z}\,}_{\!\!-}\Upsilon^\dagger(\bz))\cdot\nabla\widehat{H}(\bz)
,
\]
which arises from the relation $(\,\widehat{\!\boldsymbol{\cal Z}\,}_{\!\!+}\Upsilon(\bz))^\dagger=\,\widehat{\!\boldsymbol{\cal Z}\,}_{\!\!-}\Upsilon^\dagger(\bz)$. At this point, we restrict to finite dimensions and, upon taking the time derivative of the definition \eqref{Ddef}, one obtains
\begin{align*}
\partial_t{\widehat{\cal D}}=&\
%\frac{\de}{\de t}\left[\Upsilon_\alpha\Upsilon_\beta^*
%+
%\operatorname{\sf div}[\Upsilon_\alpha\widehat{\cal Z}_-\Upsilon_\beta^*)\right)
%\\
%=&\ 
\frac{\de}{\de t}\left(\Upsilon\Upsilon^\dagger
-
\operatorname{\sf div}\left(J\boldsymbol{\cal A}\,\Upsilon\Upsilon^\dagger\right)
+i\hbar\{\Upsilon,\Upsilon^\dagger\}
\right)
\\
=&\
-i\hbar^{-1}\big(\widehat{H}+
J\boldsymbol{\cal A}\cdot\nabla\widehat{H}\big)\Upsilon\Upsilon^\dagger
+\{\widehat{H},\Upsilon\}\Upsilon^\dagger
\\&\ 
+
i\hbar^{-1}\Upsilon\Upsilon^\dagger\big(\widehat{H}+
J\boldsymbol{\cal A}\cdot\nabla\widehat{H}\big)-\Upsilon\{\Upsilon^\dagger,\widehat{H}\}
\\&\ 
-\operatorname{\sf div}\left(J\boldsymbol{\cal A}
\left(-i\hbar^{-1}\big(\widehat{H}
+J\boldsymbol{\cal A}\cdot\nabla\widehat{H}\big)\Upsilon\Upsilon^\dagger
+\{\widehat{H},\Upsilon\}\Upsilon^\dagger
\right)
\right)
\\&\ 
-\operatorname{\sf div}\left(J\boldsymbol{\cal A}
\left(
i\hbar^{-1}\Upsilon\Upsilon^\dagger\big(\widehat{H}
+J\boldsymbol{\cal A}\cdot\nabla\widehat{H}\big)
-\Upsilon\{\Upsilon^\dagger,\widehat{H}\}
\right)
\right)
\\&\ +i\hbar
\left\{\left(-i\hbar^{-1}\big(\widehat{H}
+J\boldsymbol{\cal A}\cdot\nabla\widehat{H}\big)\Upsilon
+\{\widehat{H},\Upsilon\}\right),\Upsilon^\dagger\right\}
\\&\ +i\hbar
\left\{
\Upsilon
,\left(
i\hbar^{-1}\Upsilon^\dagger\big(\widehat{H}
+J\boldsymbol{\cal A}\cdot\nabla\widehat{H}\big)
-\{\Upsilon^\dagger,\widehat{H}\}
\right)
\right\}
\end{align*}
We recall that in the present notation all quantities are evaluated at $\bz$, e.g. $\Upsilon\Upsilon^\dagger$ stands for $\Upsilon(\bz)\Upsilon^\dagger(\bz)$. 
We expand the divergence $\operatorname{\sf div}\!\big[\widehat{H}+{\nabla\widehat{H}\cdot J\boldsymbol{\cal A}},\Upsilon\Upsilon^\dagger J\boldsymbol{\cal A}\big]\!$ in the 4th and 5th lines and we use the Leibniz product rule and the Jacobi identity in the last two lines. Then, a few cancelations yield
\begin{align*}
\partial_t{\widehat{\cal D}}_{\alpha\beta}=&\
-i\hbar^{-1}\big[\widehat{H},\Upsilon\Upsilon^\dagger-\operatorname{\sf div}\!\big(J\boldsymbol{\cal A}\Upsilon\Upsilon^\dagger\big)\big]_{\alpha\beta}
+\big\{
\widehat{H},\Upsilon\Upsilon^\dagger
\big\}_{\alpha\beta}
-
\big\{
\Upsilon\Upsilon^\dagger,\widehat{H}
\big\}_{\alpha\beta}
\\
&\ +
\operatorname{\sf div}\!\Big(
\big\{J\boldsymbol{\cal A}\Upsilon\Upsilon^\dagger,\widehat{H}\big\}
-
\big\{\widehat{H},J\boldsymbol{\cal A}\Upsilon\Upsilon^\dagger\big\}
+i\hbar^{-1}\big[J\boldsymbol{\cal A}\cdot\nabla \widehat{H},J\boldsymbol{\cal A}\Upsilon\Upsilon^\dagger\big]
\Big)_{\alpha\beta}
\\
&
 +
\operatorname{\sf div}\!\left(
\big\{\widehat{H}_{\alpha\gamma},J\boldsymbol{\cal A}\Upsilon_\beta^*\big\}\Upsilon_\gamma
-
\big\{J\boldsymbol{\cal A}\Upsilon_\alpha,\widehat{H}_{\gamma\beta}\big\}\Upsilon_\gamma^*
\right)
\\&\ 
+\Upsilon_\gamma\big\{J\boldsymbol{\cal A}\cdot\nabla \widehat{H}_{\alpha\gamma},\Upsilon_\beta^*\big\}
-
\big\{
\Upsilon_\alpha,
J\boldsymbol{\cal A}\cdot\nabla \widehat{H}_{\gamma_\beta}
\big\}\Upsilon_\gamma^*
\\
&\ 
+\big[\widehat{H}+J\boldsymbol{\cal A}\cdot\nabla \widehat{H}\,,\,\{\Upsilon,\Upsilon^\dagger\}\big]_{\alpha\beta}
+\{\widehat{H},i\hbar\{\Upsilon,\Upsilon^\dagger\}\}_{\alpha\beta}
-\{i\hbar\{\Upsilon,\Upsilon^\dagger\},\widehat{H}\}_{\alpha\beta}
\\& \ 
-i\hbar\{\Upsilon_\gamma,\{\widehat{H}_{\alpha\gamma},\Upsilon^*_{\beta}\}\}
+
i\hbar\{\{\Upsilon_\alpha,\widehat{H}_{\gamma\beta}\},\Upsilon^*_\gamma\}
\,.
\end{align*}
Then, expanding the first two divergences in the second line yields equation \eqref{Deqn}.

At this stage, we can verify the relations \eqref{PartialTraces} explicitly. We begin by proving the first in \eqref{PartialTraces}, that is by computing $\int\!\partial_t\widehat{\cal D} \,\de^6z$. This is easily done  by using the relation
\begin{equation*}
\int\{J\boldsymbol{\cal A}\cdot\nabla\widehat{H}_{\alpha\gamma},\Upsilon_\beta^*\}\Upsilon_\gamma\,\de^6z
-
\!\int
\{\Upsilon_\alpha,{J\boldsymbol{\cal A}\cdot\nabla\widehat{H}_{\gamma\beta}}\}\Upsilon_\gamma^* \,\de^6z
=-
\int\!\big[J\boldsymbol{\cal A}\cdot\nabla\widehat{H} ,\{\Upsilon,\Upsilon^\dagger\}\big]_{\alpha\beta}\,\de^6z
\,,
\end{equation*}
which indeed yields the first in \eqref{PartialTraces}.
Analogously, the second in \eqref{PartialTraces} is recovered by computing $\operatorname{\sf Tr}\partial_t\widehat{\mathcal{D}}$. The trace of the terms on the right hand side of \eqref{Deqn} are obtained as follows:
\begin{align}\nonumber
&\operatorname{\sf Tr}\big[\big\{
\widehat{H},\widehat{\cal D}
\big\}
-
\big\{
\widehat{\cal D},\widehat{H}
\big\}\big]
=2\operatorname{\sf Tr}\big\{
\widehat{H},\widehat{\cal D}
\big\}\,,
\\
\label{comput2}
&\operatorname{\sf Tr}\big[
\big\{J\boldsymbol{\cal A}\Upsilon\Upsilon^\dagger,\nabla\widehat{H}\big\}
-
\big\{\nabla\widehat{H},J\boldsymbol{\cal A}\Upsilon\Upsilon^\dagger\big\}
\big]= -2\operatorname{\sf Tr}\big\{\nabla\widehat{H}, J\boldsymbol{\cal A}\Upsilon\Upsilon^\dagger\big\}\,,
\\
\nonumber
&
\operatorname{\sf div}\!\big(
\big\{\widehat{H}_{\alpha\gamma},J\boldsymbol{\cal A}\Upsilon_\alpha^*\big\}\Upsilon_\gamma
-
\big\{J\boldsymbol{\cal A}\Upsilon_\alpha,\widehat{H}_{\gamma\alpha}\big\}\Upsilon_\gamma^*
\big)
\\
&\hspace{5cm} =
\operatorname{\sf div}\!\big[\operatorname{\sf Tr}\big\{\widehat{H}, J\boldsymbol{\cal A}\Upsilon\Upsilon^\dagger\big\}+ \operatorname{\sf Tr}\big(\big\{\widehat{H}, J\boldsymbol{\cal A}\big\}\Upsilon\Upsilon^\dagger\big)\big],
\label{comput3}
\\
\label{comput4}
&\Upsilon_\gamma\{J\boldsymbol{\cal A}\cdot\nabla\widehat{H}_{\alpha\gamma},\Upsilon_\alpha^*\}
-
\{\Upsilon_\alpha,J\boldsymbol{\cal A}\cdot\nabla\widehat{H}_{\gamma\alpha}\}\Upsilon_\gamma^*
=\operatorname{\sf Tr}\{J\boldsymbol{\cal A}\cdot\nabla\widehat{H},\Upsilon\Upsilon^\dagger\}\,,
\end{align}
as well as
\begin{equation}\label{comput5}
-i\hbar\{\Upsilon_\gamma,\{\widehat{H}_{\alpha\gamma},\Upsilon^*_{\alpha}\}\}
+
i\hbar\{\{\Upsilon_\alpha,\widehat{H}_{\gamma\alpha}\},\Upsilon^*_\gamma\}=-i\hbar\operatorname{\sf Tr}\big\{\widehat{H}, \big\{\Upsilon, \Upsilon^\dagger\big\}\big\}.
\end{equation}
The trace of the other terms vanishes.
After several computations, one obtains that the sum of \eqref{comput2}, \eqref{comput3}, \eqref{comput4}, and \eqref{comput5} gives
\[
\operatorname{\sf Tr}\big( \{\widehat{H}, J\boldsymbol{\cal A}\big\}\cdot \nabla (\Upsilon \Upsilon^\dagger)\big) +\operatorname{\sf Tr}\big( \nabla \widehat{H}\cdot \big\{J\boldsymbol{\cal A} ,\Upsilon \Upsilon^\dagger\big\}\big) -\operatorname{\sf Tr}\big\{\widehat{H}, i\hbar \big\{\Upsilon, \Upsilon^\dagger\big\} - \operatorname{\sf div}\big(J\boldsymbol{\cal A} \Upsilon\Upsilon^\dagger\big)\big\}.
\]
Hence the second equation in \eqref{PartialTraces} follows if the following identity holds:
\[
\operatorname{\sf Tr}\big( \{\widehat{H}, J\boldsymbol{\cal A}\big\}\cdot \nabla (\Upsilon \Upsilon^\dagger)\big) + \operatorname{\sf Tr}\big( \nabla \widehat{H}\cdot \big\{J\boldsymbol{\cal A} ,\Upsilon \Upsilon^\dagger\big\}\big)= -\operatorname{\sf Tr}\big\{\widehat{H}, \Upsilon \Upsilon^\dagger\big\},
\]
for all $\widehat{H}$ and $\Upsilon$. This identity is equivalent to
$J \big(\nabla \boldsymbol{\cal A} - (\nabla \boldsymbol{\cal A})^T\big) J=J$,
which follows since $\nabla\boldsymbol{\cal A}-(\nabla\boldsymbol{\cal A})^T=-J$ and $J^2=-1$.


\bigskip

\begin{thebibliography}{10}\small

\smallskip

\bibitem{Aleksandrov}
Aleksandrov, I.V. {\it The statistical dynamics of a system consisting of a classical and a quantum subsystem}. {Z. Naturforsch.} 36a (1981), 902-908

\bibitem{Anderson}
Anderson, A. {\it Quantum backreaction on ``classical'' variables}. {Phys. Rev. Lett.} 74 (1995), 621-625

\bibitem{Barcelo}
Barcel\'o, C.; Carballo-Rubio, R.; Garay, L.J.; G\'omez-Escalante, R. {\it Hybrid classical-quantum formulations ask for hybrid notions}. {Phys. Rev. A} 86 (2012), 042120 

\bibitem{BiMo}
Bialynicki-Birula, I.; Morrison, P.J. {\it Quantum mechanics as a generalization of Nambu dynamics to the Weyl-Wigner formalism}. {Phys. Lett. A} 158 (1991), 453-457

\bibitem{Bohr1935}
Bohr, N. {\it Can quantum-mechanical description of physical reality be considered complete?}. {Phys. Rev.} 48 (1935), 696-702


\bibitem{Bondar}
Bondar, D.; Cabrera, R., Lompay, R.R.; Ivanov, M.Yu.; Rabitz, H. {\it Operational dynamic modeling transcending quantum and classical mechanics}. {Phys. Rev. Lett.} 109 (2012), 190403 

\bibitem{BLTr1}
Bonet-Luz, E.; Tronci, C. {\it Geometry and symmetry of quantum and classical-quantum variational principles}.  {J. Math. Phys.} 56 (2015), 082104 

\bibitem{BLTr2}
Bonet-Luz, E.; Tronci, C. {\it Hamiltonian approach to Ehrenfest expectation values and Gaussian quantum states}. {Proc. R. Soc. A} 472 (2016), 20150777

\bibitem{boucher}
Boucher, W.; Traschen, J. {\it Semiclassical physics and quantum fluctuations}. {Phys. Rev. D} 37 (1988), 3522-3532

\bibitem{Mezic}
Budi\v{s}i\'{c}, N.; Mohr, R.; Mezi\'{c}, I. {\it Applied Koopmanism}. Chaos 22 (2012), 047510.

\bibitem{CaroSalcedo}
Caro, J.; Salcedo, L.L {\it Impediments to mixing classical and quantum dynamics}. {Phys. Rev. A} 60 (1999), 842-852

\bibitem{ChernoffMarsden}
Chernoff, P.R.; Marsden, J.E. {\it Some remarks on Hamiltonian systems and quantum mechanics}. In ``Foundations of Probability Theory, Statistical Inference, and Statistical Theories of Science'' Harper and Hooker (eds.) Vol. III, 35-53.  Reidel Publishing Company, Dordrecht-Holland. 1976

\bibitem{Marmo}
Chru\'sci\'nski, D.;  Kossakowski, A.; Marmo, G.; Sudarshan, E.C.G. {\it Dynamics of interacting classical and quantum systems}. {Open. Syst. Inf. Dyn.} 18 (2011), 339 

\bibitem{Clebsch}
Clebsch, A. {\it \"Uber die Integration der hydrodynamischen Gleichungen}. {J. Reine Angew. Math.} 56 (1859). 1-10

\bibitem{deGosson}
de Gosson, M.A. {\it Symplectically covariant Schr\"odinger equation in phase space}. {J. Phys. A: Math. Gen.} 38 (2005), 9263-9287

\bibitem{Wiener}
Della Riccia, G.; Wiener, N. {\it Wave mechanics in classical phase space, Brownian motion, and quantum theory}. {J. Math. Phys.} 6 (1966), 1372-1383

\bibitem{Diosi}
Di\'osi, L.; Halliwell, J.J. {\it Coupling classical and quantum variables using continuous quantum measurement theory}. {Phys. Rev. Lett.} 81 (1998), 2846

\bibitem{Elze}
Elze, H.T. {\it Four questions for quantum-classical hybrid theory}. {J. Phys.: Conf. Ser.} 361 (2012) 012004

\bibitem{Feynman}
Feynman, R. ``Feynman's Thesis $-$ A New Approach to Quantum Theory''. Edited by L.M. Brown. World Scientific (2005)

\bibitem{FeynmanNegativeProb}
Feynman, R.  {\it Negative probability}. {In}  ``Quantum Implications: Essays in Honour of David Bohm''.  Hiley and Peat (eds.) 235--248.  Routledge \& Kegan Paul Ltd. 1987.

\bibitem{GBTr19}
Gay-Balmaz, F.; Tronci, C. {\it Madelung transform and probability currents in hybrid classical-quantum dynamics
}. In preparation.


\bibitem{GBTr}
Gay-Balmaz, F.; Tronci, C. {\it Vlasov moment flows and geodesics on the Jacobi group}. J. Math. Phys. 53 (2012), 123502 

\bibitem{Gerasimenko}
Gerasimenko, V. {\it Dynamical equations of quantum-classical systems}. {Theor. Math. Phys.} 50 (1982), 49-55 

\bibitem{Ghose}
Ghose, P. {\it The unfinished search for wave-particle and classical-quantum harmony}. {J. Adv. Phys.} 4 (2015), 236-251

\bibitem{GoKoSu}
Gorini, V.; Kossakowski, A.; Sudarshan, E.C.G. {\it Completely positive semigroups of $N-$level systems}. {J. Math. Phys.} 17 (1976). 821

\bibitem{Sternberg}
Guillemin, V; Sternberg, S. {\it The moment map and collective motion}. {Ann. Phys.} 27 (1980), 220-253 

\bibitem{Sternberg2}
Guillemin, V; Sternberg, S. Symplectic Techniques in Physics. Cambridge University Press. 1984.

\bibitem{Gunther}{
G\"unther, P. {\it Presymplectic manifolds and the quantization of relativistic particle systems}. In ``Differential geometrical methods in mathematical physics'' (Proc. Conf., Aix-en-Provence/Salamanca, 1979), 383–400, Lecture Notes in Math. 836, Springer, 1980}

\bibitem{Hall}
Hall, B.C. Quantum Theory for Mathematicians. Springer. 2013

\bibitem{HaRe}
Hall, M.J.W.; Reginatto, M. {\it Interacting classical and quantum ensembles}. {Phys. Rev. A} 72 (2005), 062109

\bibitem{HacohenGourgy}
Hacohen-Gourgy, S.; Martin, L.S.; Flurin, E.; Ramasesh, V.V.; Whaley, K.B.; Siddiqi, I. {\it Quantum dynamics of simultaneously measured non-commuting observables}. {Nature} 538 (2016), 491-494

\bibitem{HoKu}
Holm, D.D.; Kupershmidt, B.A. {\it Poisson brackets and Clebsch representations for magnetohydrodynamics, multifluid plasmas, and elasticity}. {Phys. D} 6 (1983), 347-363

\bibitem{HoScSt}
Holm, D.D.; Schmah, T.; Stoica, C. Geometric Mechanics and Symmetry: From Finite to Infinite Dimensions. OUP. 2009

%\bibitem{Jackiw}
%Jackiw, R. %{MIT report CTP-2516}. 
%{\tt arXiv:hep-th/9604040} (1996)

\bibitem{Sugny}
Jauslin, H.R.; Sugny, D. {\it Dynamics of mixed classical-quantum systems, geometric quantization and coherent states}. In ``Mathematical Horizons for Quantum Physics''.  Lect. Notes Ser. Inst. Math. Sci. Natl. Univ. Singap., 20, 65-96. World Scientific. 2010.

\bibitem{Kapral}
Kapral, R. {\it Progress in the theory of mixed quantum-classical dynamics}. {Annu. Rev. Phys. Chem.} 57 (2006), 129-57

\bibitem{Khrennikov}
Khrennikov,  A. Non-Archimedean Analysis: Quantum Paradoxes, Dynamical Systems and Biological Models. Kluwer Academic Publishers. 1997.

\bibitem{Kirillov}
Kirillov, A. A. {\it Geometric quantization}. In ``Dynamical Systems IV'', 139-176, Encyclopaedia Math. Sci., 4, Springer, 2001.

\bibitem{Klein}
Klein, U. {\it From Koopman-von Neumann theory to quantum theory}. {Quantum Stud.: Math. Found.} 5 (2018), 219-227

\bibitem{Koopman}
 Koopman, B.O. {\it Hamiltonian systems and transformations in Hilbert space}. {Proc. Nat. Acad. Sci.} 17 (1931), 315
 
 \bibitem{Kostant1}{
Kostant, B. {\it Line bundles and the prequantized Schr\"odinger equation}. In ``Colloquium on Group Theoretical Methods in Physics''. Centre de Physique Th\'eorique, Marseille, June 1972, IV.1-IV.22}
 
\bibitem{Kostant}
Kostant, B. {\it Quantization and unitary representations}, In ``Lectures in modern analysis and applications III'', 87--208. Lecture Notes in Math. 170, Springer, 1970

\bibitem{Lindblad}
Lindblad, G. {\it On the generators of quantum dynamical semigroups}. {Commun. Math. Phys.} 48 (1976), 119

\bibitem{Lindblad2}
Lindblad, G. {\it Completely positive maps and entropy inequalities}. {Commun. Math. Phys.} 40 (1975), n. 2, 147-151

\bibitem{MaRa}
Marsden, J.E.; Ratiu, T.S. Introduction to Mechanics and Symmetry. Springer. 1998

\bibitem{MaWe}
Marsden, J.E.; Weinstein, A. {\it Coadjoint orbits, vortices, and Clebsch variables for incompressible fluids}. {Phys. D} 7 (1983), 305-323

\bibitem{Mauro}
Mauro, D. {\it On Koopman-von Neumann waves}. {Int. J. Mod. Phys. A} 17 (2002), 1301

\bibitem{Morrison}
Morrison, P.J. {\it Hamiltonian description of the ideal fluid}. {Rev. Mod. Phys.} 70 (1998), 467-521

\bibitem{Ciccotti}
Nielsen, S.; Kapral R.; Ciccotti, G. {\it Statistical mechanics of quantum-classical systems}. {J. Chem. Phys.} 115 (2001), 5805--5815

\bibitem{PeTe}
Peres, A.; Terno, D.R. {\it Hybrid classical-quantum dynamics}. {Phys. Rev. A} 63 (2001), 022101 

\bibitem{Prezhdo}
Prezhdo, O.V. {\it A quantum-classical bracket that satisfies the Jacobi identity}. {J. Chem. Phys.} 124 (2006), 201104 

\bibitem{PrKi}
Prezhdo, O.V.; Kisil, V.V. {\it Mixing quantum and classical mechanics}. {Phys. Rev. A} 56 (1997), 162-175

\bibitem{Buric}
Radonji\'c, M.; Prvanovi\'c, S.; Buri\'c, N. {\it Hybrid quantum-classical models as constrained quantum systems}. {Phys. Rev. A} 85 (2012), 064101

\bibitem{Ramos}
Ramos-Prieto, I.; Urz\'ua-Pineda, A.R.; Soto-Eguibar, F.; Moya-Cessa, H.M. 
{\it KvN mechanics approach to the time-dependent frequency harmonic oscillator}. 
{Sci. Rep.} 8 (2018), 8401

\bibitem{Sahoo}
Sahoo, D. {\it Mixing quantum and classical mechanics and uniqueness of Planck's constant}. {J. Phys. A: Math. Gen.} 37 (2004), 997-1010

\bibitem{Salcedo}
Salcedo, L.L. {\it Absence of classical and quantum mixing}. Phys. Rev. A 54 (1996), n. 4, 3657-3660

\bibitem{Scully1}
Schleich, W.P.; Greenberger, D.M.; Kobe, D.H.; Scully, M.O. {\it Schr\"odinger equation revisited}. {PNAS} 110 (2013), 5374-5379

\bibitem{Scully2}
Schleich, W.P.; Greenberger, D.M.; Kobe, D.H.; Scully, M.O. {\it A wave equation interpolating between
classical and quantum mechanics}. {Phys. Scr.} 90 (2015), 108009

\bibitem{Sergi}
%Sergi, A. {\it Non-Hamiltonian commutators in quantum mechanics}. {Phys. Rev. E} 72 (2005), 066125
{
Sergi, A.; Hanna, G.; Grimaudo, R.; Messina, A. {\it Quasi-Lie brackets and the breaking of time-translation symmetry for quantum systems embedded in classical baths}. \makebox{Symmetry 10 (2018),  518}}

\bibitem{Shirokov}
Shirokov, Yu.M. {\it Quantum and classical mechanics in the phase space representation}. {Sov. J. Part. Nucl.} 10 (1979), 1-18




\bibitem{Souriau}{
Souriau, J.-M. {\it Quantification g\'eom\'etrique}, {Comm. Math. Phys.} 1 (1966), 374-398
}

\bibitem{Sudarshan}
Sudarshan, E.C.G. {\it Interaction between classical and quantum systems and the measurement of quantum observables}. {Pr\={a}ma\d{n}a} 6 (1976), 117-126

\bibitem{Sudarshan2}
Sudarshan, E.C.G. {\it Consistent measurement of a quantum dynamical variable using classical apparatus}. \href{https://arxiv.org/abs/quant-ph/0402134}{\tt arXiv:quant-ph/0402134} (2004)

\bibitem{Terno}
Terno, D.R. {\it Inconsistency of quantum—classical dynamics, and what it implies}. {Found. Phys.} 36 (2006), 102-111

\bibitem{tHooft}
't Hooft, G. {\it Quantummechanical behaviour in a deterministic model}. {Found. Phys. Lett.} 10 (1997), 105-111

\bibitem{Tronci18}
Tronci, C. {\it Momentum maps for mixed states in quantum and classical mechanics}. J. Geom. Mech. (2019). To appear. \href{https://arxiv.org/abs/1810.01332}{\tt  arXiv:1810.01332} 


\bibitem{VanHove}
van Hove, L. On certain unitary representations of an infinite group of transformations. PhD Thesis (1951). Word Scientific 2001

\bibitem{Viennot}
Viennot, D.; Aubourg, L. 
{\it Schr\"odinger?Koopman quasienergy states of quantum systems driven by classical flow}. 
{J. Phys, A: Math.  Theor.}  51 (2018), 335201 


\bibitem{VonNeumann}
von Neumann, J. Mathematical Foundations of Quantum Mechanics. Princeton University Press, 1955

\bibitem{VonNeumann2}
von Neumann, J. {\it Zur Operatorenmethode In Der Klassischen Mechanik }. {Ann. Math.} 33 (1932). 587-642

\bibitem{Wilczek}
Wilczek, F. {\it Notes on Koopman von Neumann mechanics, and a step beyond}. { Unpublished}. 2015

\bibitem{Zurek}
Zurek, W.H. {\it Decoherence, einselection, and the quantum origins of the classical}. {Rev. Mod. Phys.} 75 (2003), 705

\bibitem{Material}
All the source codes used to arrive at the results (including plotting) of this section can be found at \url{https://github.com/dibondar/QCHybrid}.

\end{thebibliography}
\end{document}